\documentclass[
  twocolumn,
  prb,
  showpacs,
  amsmath,
  amssymb,
  superscriptaddress,
  floatfix,
 ]{revtex4}

\usepackage{bm}
\usepackage{bbm}
\usepackage{graphicx}
\newcommand{\Real}{\mathop{\rm Re}}
\begin{document}

\title{Luttinger liquids with multiple Fermi edges: Generalized
  Fisher-Hartwig conjecture and numerical analysis
  of Toeplitz determinants} 

\author{I.\ V.\ Protopopov}
\affiliation{
 Institut f\"ur Nanotechnologie, Karlsruhe Institute of Technology,
 76021 Karlsruhe, Germany
}
\affiliation{
 L.\ D.\ Landau Institute for Theoretical Physics RAS,
 119334 Moscow, Russia
}

\author{D.\ B.\ Gutman}
\affiliation{Department of Physics, Bar Ilan University, Ramat Gan 52900,
Israel }

\author{A.\ D.\ Mirlin}
\affiliation{
 Institut f\"ur Nanotechnologie, Karlsruhe Institute of Technology,
 76021 Karlsruhe, Germany
}
\affiliation{
 Institut f\"ur Theorie der kondensierten Materie and DFG Center for
 Functional Nanostructures, 
 Karlsruhe Institute of Technology, 76128 Karlsruhe, Germany
}
\affiliation{
 Petersburg Nuclear Physics Institute,
 188300 St.~Petersburg, Russia
}

\begin{abstract}
It has been shown that solutions of a number of many-body problems out
of equilibrium can be expressed in terms of Toeplitz determinants with
Fisher-Hartwig (FH) singularities. In
the present paper, such Toeplitz determinants are studied numerically.
Results of our numerical calculations  fully agree 
with the FH conjecture in an extended form 
that includes a summation over all FH representations (corresponding to
different branches of the logarithms). 
As specific applications,  we consider problems of Fermi edge singularity and 
tunneling spectroscopy of Luttinger liquid with multiple-step energy
distribution functions, including the case of population inversion.
In the energy representation, a sum over FH branches produces 
power-law singularities at multiple edges.

\end{abstract}


\maketitle

\section{Introduction}
For more than half a century, quantum many-body systems remain one of central research
directions in the condensed matter physics.   
There is a number of quantum many-body problems that are of fundamental
physical importance and, at the same time, possess an exact solution. 
These are the Anderson orthogonality catastrophe\cite{Anderson}, Fermi edge
singularity\cite{Nozieres} (FES), Luttinger liquid\cite{Tomonaga} (LL)
zero-bias anomaly\cite{Luther}, and Kondo problems\cite{Kondo}.  
It has been realized long ago   that  these problems  are,  in fact,
deeply interconnected,   both from the point of view  of the  underlying physics 
and of the  mathematics  involved. Such connections have been used, e.g.,  for the
representation of  the dynamics of the  Kondo problem  
as  an infinite sequence of  Fermi-edge-singularity events\cite{Yuval}. 
These relationships between many-body problems extends beoynd fermions and encompass 
also interacting bosons (e.g., the Lieb-Liniger model\cite{Lieb_Liniger}), 
one-dimensional Heisenberg chains and etc.\cite{Korepin}

In recent works by two of us with
Gefen\cite{GGM_long2010,GGM_short2010,Gutman10},   
non-equilibrium realizations of some of these problems have been
investigated. For this purpose, 
we  have developed a non-equilibrium bosonization technique
generalizing the conventional bosonization
\cite{stone,Delft,Gogolin,giamarchi,maslov-lectures} onto problems
with non-equilibrium distribution functions.
We have shown that the relevant correlation functions  can be
expressed through  Fredholm determinants of ``counting'' operator. 
The information on the specific type of the problem, as well as on different 
aspects of the interaction, 
is  encoded  in the time-dependent scattering phase of the counting operator.
The findings of 
Refs.~\onlinecite{GGM_long2010,GGM_short2010,Gutman10} 
have demonstrated  that
the above classical many-body problems are even more closely connected than has been 
previously understood, extending the interrelations into the
non-equilibrium regime.  

The ``counting'' operators governing the simplest (one-particle)
non-equilibrium Green 
functions in the above models can be reduced to Toeplitz matrix form upon
regularization and discretization \cite{Gutman10}. The electron
energy distribution function then determines the symbol of the
Toeplitz matrix. The most
interesting situation arises when the distribution function has
multiple steps (``Fermi edges''), which results in step-like 
singularities of the symbol. According to the Fisher-Hartwig
conjecture \cite{fisher-hartwig}, this 
leads to a non-trivial power-law behavior
of the correlation functions. Recent progress in the analysis of
Toeplitz determinants with Fisher-Hartwig singularities has allowed to
establish their leading asymptotic behavior \cite{deift09}. In
Ref.~\onlinecite{Gutman10}   
a generalized Fisher-Hartwig conjecture was put forward that includes
a summation over all Fisher-Hartwig representations (corresponding to
different branches of the logarithm of the symbol). This yields   
also terms with subleading power-law factors. While these terms are
formally smaller (as compared to the leading term) when one considers
the Green function in the time representation, they contain different
oscillatory exponents. Therefore, after a transformation to 
the energy representation, they produce power-law singularities at
different edges, which makes these terms physically important. 
The extended version of the Fisher-Hartwig conjecture is also expected
to be of interest from the purely mathematical point of view. 

In the present paper we perform a numerical analysis of Toeplitz determinants 
with Fisher-Hartwig singularities.
The results of numerical calculation  fully confirm the extended
conjecture for the asymptotic (long-time or low-energy)
behavior. Furthermore, the numerics allows us 
to explore correlation functions in the entire energy range.
To be specific, we focus on two fermionic problems: (i) the
Fermi-edge singularity in X-ray absorption and (ii) the tunneling
density of states (TDOS) of a non-equilibrium Luttinger liquid. 

The structure of the present paper is as follows.
Section \ref{Many-particle-problems} contains a brief review of the
connection between  
one-particle correlation functions of many-body problems and Toeplitz determinants.
In Sec.~\ref{Sec:Asymptotic_properties_of_Toeplitz_determinants}
we present the extended version of the Fisher-Hartwig conjecture,
as well as illustrate it and discuss its implucations
on examples relevant to our many-body problems. 
In Sec.~\ref{Sec:NumericalAnalysis} 
we calculate the Toeplitz determinants (and thus the correlation
functions under interest) numerically   and compare the exact results
with the asymptotic formulas.  
Our findings are summarized in Section \ref{Sec:Conclusions}, where we
also discuss prospects for future research.

\section{Many-particle problems as Fredholm determinants}
\label{Many-particle-problems}

\subsection{Fermi edge singularity}

The FES problem describes the  scattering of conduction  electrons off a
localized hole which  is left behind by  an electron excited into the
conduction band. 
Historically, the FES problem was first solved by exact summation of an
infinite diagrammatic series \cite{Nozieres}.
While in the FES problem there is no interaction between electrons
in the conducting band, it has many features characteristic of genuine
many-body physics. 
Despite the fact that conventional experimental realizations of FES are
three-dimensional, the problem can be reduced (due to the local and isotropic 
character of the interaction with the core hole) to that of
one-dimensional chiral fermions. For this reason,  bosonization
technique can be effectively applied, leading
to  an alternative and very elegant solution \cite{Schotte}.

\begin{figure}
\includegraphics[width=190pt]{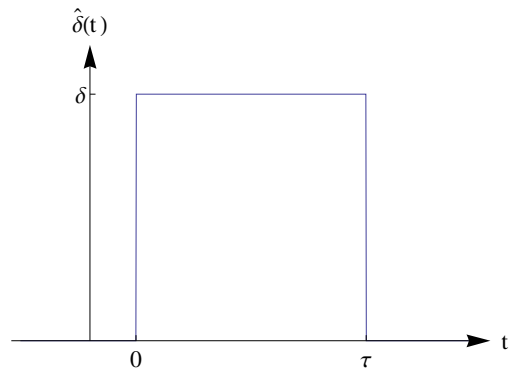}
\caption{\small Rectangular shaped pulse in the phase $\delta(t)$. }
\label{deltaPulse}
\end{figure}

One can consider the FES out of equilibrium\cite{Abanin}, 
with an arbitrary electron distribution function $n(\epsilon)$. 
This problem can be solved within the framework 
of non-equilibrium bosonization\cite{GGM_long2010}, with the following
results for the  emission/absorption rates: 
\begin{equation}
\label{b1}
iG^\gtrless_{\rm FES}(\tau)=\pm
{ \Lambda{\overline{\Delta}}_{\tau}(2\pi-2\delta_0) \over
2\pi v (1\pm i\Lambda\tau)^{(1-\delta_0/\pi)^2}}\,.
\end{equation}
Here $\delta_0$is the $s$-wave electronic phase shift  due to the
scattering of conduction electrons off the core hole. Further,
$\overline{\Delta}_\tau[2\pi-2\delta_0]$ is the Fredholm determinant 
(normalized to its value at zero temperature) 
\begin{equation}
 \overline{\Delta}_{\tau}[\delta]\equiv\frac{\Delta_\tau[\delta]}{\Delta_{\tau}[\delta,
     T=0]}=\frac{\det[1+(e^{-i\hat{\delta}}-1)\hat{n}]}
{\det[1+(e^{-i\hat{\delta}}-1)\hat{n}\left(T=0\right)]}\,.     
\label{DetBasic}
\end{equation}
The phase $\hat{\delta}$  is an operator local in time $t$ conjugate
to electron energy $\epsilon$ and has  
characteristic rectangular shape (Fig. \ref{deltaPulse})
\begin{equation}
 \hat{\delta}(t)=\delta \left[\theta(t)-\theta(t-\tau)\right]\,.
\label{delta_t}
\end{equation}
The connection of the non-equilibrium FES problem to Fredholm
determinants is summarized in the first row of Table 
\ref{table1}. 

\subsection{Luttinger liquid: tunneling spectroscopy}

The tunneling spectroscopy technique allows one to explore
experimentally Keldysh
Green functions of an interacting system that carry information about
both tunneling density of states and energy distribution. 
Recent experiments on carbon nanotubes and quantum Hall edges have
proved the efficiency of this technique in the context of 1D
systems\cite{tunnel-spectroscopy,tunnel-spectroscopy-qhe}. The
technological and experimental 
advances motivate the theoretical interest in the tunneling
spectroscopy of strongly correlated 1D structures away from
equilibrium
\cite{GGM_long2010,jakobs08,gutman08,trushin08,pugnetti09,Ngo,takei10,bena10}.

In the case of a  LL formed by 1D interacting fermions, the Keldysh
Green function may be 
evaluated theoretically via the non-equilibrium bosonization technique. 
Assuming that a long LL conductor is adiabatically  coupled to
two reservoirs (modeled as non-interacting 1D
wires\cite{Maslov,Safi,Ponomarenko}) with 
distribution functions $n_R(\epsilon)$ and  $n_L(\epsilon)$ respectively, 
one obtains for the Green functions of the right movers
\cite{GGM_long2010}   
\begin{equation}
\label{d3}
G^\gtrless_R(\tau)=\mp \frac{i\Lambda}{2\pi u}
\frac{\overline{\Delta}_{R\tau}[\delta_R]\overline{\Delta}_{L\tau}[\delta_L]}
{(1\pm i\Lambda \tau)^{1+\gamma}}\,,
\end{equation}
where $u=v/K$ is the sound velocity,
\begin{equation}
\gamma=(1-K)^2/2K\,,
\end{equation}
and
\begin{equation}
K=(1+g/\pi v)^{-1/2}
\end{equation}
 is the standard LL parameter in the interacting region.
The determinants $\overline{\Delta}_{\eta\tau}[\delta_\eta]$ ($\eta=R,
L$) are given by Eq. (\ref{DetBasic}) with $n(\epsilon)$ replaced by
the corresponding distribution functions $n_{\eta}(\epsilon)$ and 
\begin{equation}
\label{delta}
 \delta=\delta_{\eta}=\pi \frac{1+\eta K}{\sqrt{K}}\,.
\end{equation}
The connection of the Luttinger liquid Green functions  to Fredholm
determinants is summarized in the last two rows of Table
\ref{table1}. 

It is worth emphasizing that the the rectangular shape (\ref{delta_t})
of the pulse with the amplitude (\ref{delta}) is valid 
in the case when the coupling to reservoirs is smooth on the scale of
the plasmon wave length $v/T$, $u/T$. In the opposite regime the pulse
$\delta(t)$ entering (\ref{d3}) is fractionalized in a sequence of
rectangular pulses \cite{GGM_long2010}. In the long-wire limit the corresponding
determinant splits into a product of single-pulse  (i.e Toeplitz-type)
determinants. For definiteness, we focus on the adiabatic case in this paper.

\setlength{\tabcolsep}{7pt}
\begin{table}
\begin{center}
\resizebox{8cm}{!}{
  \begin{tabular}{ | c  |  c | c | c|}
    \hline 
       & $\delta_R$ & $\delta_L$ & $\gamma$ \\ \hline \hline
    $G_{\rm FES}^>(\tau)$ &  $2(\pi -\delta_0)$ & $0$ &
    $\frac{\delta_0^2}{\pi^2}-\frac{2\delta_0}{\pi}$\\ \hline 
    $G_{R}^>(\tau)$    &  $2\pi \frac{1+K}{2\sqrt{K}}$   &
    $2\pi\frac{1-K}{2\sqrt{K}}$ & $\frac{(1-K)^2}{2K}$\\ \hline  
   $G_{L}^>(\tau)$   & $2\pi\frac{1-K}{2\sqrt{K}}$ &
   $2\pi\frac{1+K}{2\sqrt{K}}$ & $\frac{(1-K)^2}{2K}$\\
\hline
 \end{tabular}
} 
\caption{Non-equilibrium correlation functions of many-body
  problems: Fermi edge singularity ($G^>_{\rm FES}(\tau)$),  Green functions
  of right- and left-moving fermions in a LL ($G^>_{\rm R}(\tau)$ and
  $G^>_{\rm L}(\tau)$). 
Evaluation of these correlation functions
  yields the results in the form of Fredholm-Toeplitz determinants
  $\tau^{-\gamma-1}\overline{\Delta}_R[\delta_R]\overline{\Delta}_L[\delta_L]$. The
  corresponding phases 
  $\delta_{R,L}$ are presented in the second and third  columns. (For
  LL an adiabatic coupling to reservoirs on the scale of the
  characteristic plasmon wave length is assumed.) The
  determinants are normalized to their values at  
zero temperature. The exponent $\gamma$ governing the zero-temperature
correlation function is shown in the last column. }  
\label{table1} 
\end{center}
\end{table}

\subsection{Ultraviolet regularization and reduction to Toeplitz matrix}

Due to characteristic rectangular shape (\ref{delta_t}) of the pulses
$\delta(t)$ the Fredholm determinants  
$\Delta_\tau(\delta)$ are in fact of the Toeplitz form.
Specifically, one can write
\begin{equation}
\label{determinant_definition2}
\Delta_\tau[\delta]=
\det[1 + \hat{P}(e^{-i\delta}-1) \hat{n}\hat{P}]\,\,.
\end{equation}
Here we have defined the projection operator
\begin{eqnarray}
\hat{P} y(t)=\left\{
\begin{array}{l}
      y(t)\, , \,\,\, {\rm for}\,\,\,  t \in [0,\tau] 
      \\ \\
      0 \,, \,\,\,\,\,\,\,\,\,\,\,  {\rm  otherwise}\,.
\end{array}
\right.
\label{projection_operator}
\end{eqnarray}

The form (\ref{determinant_definition2}) is convenient for peforming
the ultraviolet regularization of the determinant
$\Delta_\tau[\delta]$. Specifically,
we  discretize the time $t$ by 
introducing an elementary time step   $\Delta t=\pi/\Lambda$, 
such that $t_j=j \Delta t$. This corresponds to restricting the energy variable
$\epsilon$ to the range $[-\Lambda,\Lambda]$. We arrive then at a
finite-dimensional determinant 
\begin{equation}
\label{Toeplitz}
\Delta_N[\delta]=\det[f(t_j-t_k)]\,, \,\,\,  0\leq j,k \leq N-1\,.
\end{equation}
Here $N=\tau\Lambda/\pi$ and $f(t_j-t_k)$ is Fourier transform of the function 
\begin{equation}
\label{f_epsilon_non_periodic}
f(\epsilon) = 1 + n(\epsilon) (e^{-i\delta} -1)\,.
\end{equation}
The matrix elements $f(t_j-t_k)$ depend on $j$ and $k$ via the
difference $j-k$ only, so that the obtained matrix is of Toeplitz type.

In order to bring Eqs.~(\ref{Toeplitz}),
(\ref{f_epsilon_non_periodic}) to the canonical form used in the
theory of Toeplitz matrices, we have to define the function
$f(\epsilon)$ on the unit circle $|z|=1$.  This is easily done by
identifying the polar angle $\theta\in [-\pi,\pi]$ parametrizing the unit circle via
$z=e^{i\theta}$ with the appropriately rescaled energy:  $\theta=\pi\epsilon/\Lambda$.
However, if this is done directly with the function
(\ref{f_epsilon_non_periodic}), a non-physical jump will arise at
$\theta=\pm\pi$. In order to eliminate it, one has to introduce an  additional
phase factor into the definition of $f(\epsilon)$: 
\begin{equation}
\label{f_epsilon_periodic}
f(\epsilon) = [1 + n(\epsilon) (e^{-i\delta} -1)] e^{-i{\delta\over
    2}{\epsilon\over \Lambda}}\,.
\end{equation}        
After the mapping to the unit circle,
$z=e^{i\pi\epsilon/\Lambda}$, this defines a function $f(z)$ (known as
the symbol of the Toeplitz matrix) that is perfectly smooth at $z=-1$. It will,
however, have discontinuities
(``Fisher-Hartwig singularities'') at the positions
$z=e^{i\pi\epsilon_j/\Lambda}$ if the distribution function
$n(\epsilon)$ has such discontinuities (``Fermi edges'') at
$\epsilon_j$.  We will be interested in the situation when there are
several (at least two) such discontinuities.

It is worth emphasizing that the regularization (\ref{Toeplitz}),
(\ref{f_epsilon_periodic})  
makes explicit the dependence of the determinant 
$\Delta_\tau(\delta)$ on the integer part of $\delta/2\pi$ (thus
making redundant the procedure of analytical continuation from $\delta\in
[-\pi,\pi]$ to larger $|\delta|$ discussed in
Ref.~\onlinecite{GGM_long2010}). This allows us to directly compute the
determinant at arbitrarily large (by absolute value) $\delta$.  

As the matrix $\{f(t_j-t_k)\}$ with $0\leq j,k \leq N-1$ is of Toeplitz
form, results concerning the large-$N$ asymptotic behavior of Toeplitz
determinats $\Delta_N$   
can be applied. We summarize them in the next section. 
Physically, the large-$N$ limit corresponds to the regime of long time
$\tau$, i.e. to infrared asymptotics of correlation functions under
interest. (In the energy representation, this translates into
low-energy behavior around singularities.) 

Furthermore, Eqs.~(\ref{Toeplitz}), (\ref{f_epsilon_periodic}) are
also very convenient for numerical evaluation of the determinant $\Delta_\tau(\delta)$,
providing us access to the full time (or, after Fourier
transformation, energy) dependence of the correlation functions.

\section{Asymptotic properties of Toeplitz determinants}
\label{Sec:Asymptotic_properties_of_Toeplitz_determinants}

Toeplitz matrices and operators were introduced by O.~Toeplitz a
century ago. Since this time, asymptotic properties 
of Toeplitz determinants have been in a focus of interest
of mathematicians, starting from the 1915 paper \cite{szego15} that
was the first research paper by G.~Szeg\"o. 
The Szeg\"o theorems  \cite{szego39} valid for a smooth
symbol yield the large-$N$ asymptotics of the determinant, which is
exponential in $N$, with an $N$-independent prefactor.
As was realized by Fisher and Hartwig \cite{fisher-hartwig}, in the
case of a symbol with singularities, the asymptotics acquires, in
addition to the exponential factor, also a power-law factor. 
Thus,  the infrared behavior of the
Toeplitz determinant (\ref{Toeplitz}) includes non-trivial power-law factors
if  the function  $f(z)$ is not 
smooth  on the unit circle. The simplest example is the
zero-temperature determinant \cite{Gutman10} 
\begin{multline}
 \Delta_N[\delta, T=0]=e^{-i\delta N/2}\left( \frac{\pi}{\Lambda
   \tau}\right)^{\left(\frac{\delta}{2\pi}\right)^2}\\\times
 G\left(1-\frac{\delta}{2\pi}\right)G\left(1+\frac{\delta}{2\pi}\right)\, 
\label{DeltN0}
\end{multline}
that has a power-law dependence on time  in the long-time limit  
($\Lambda\tau\gg 1)$.

\begin{figure*}
\includegraphics[width=190pt]{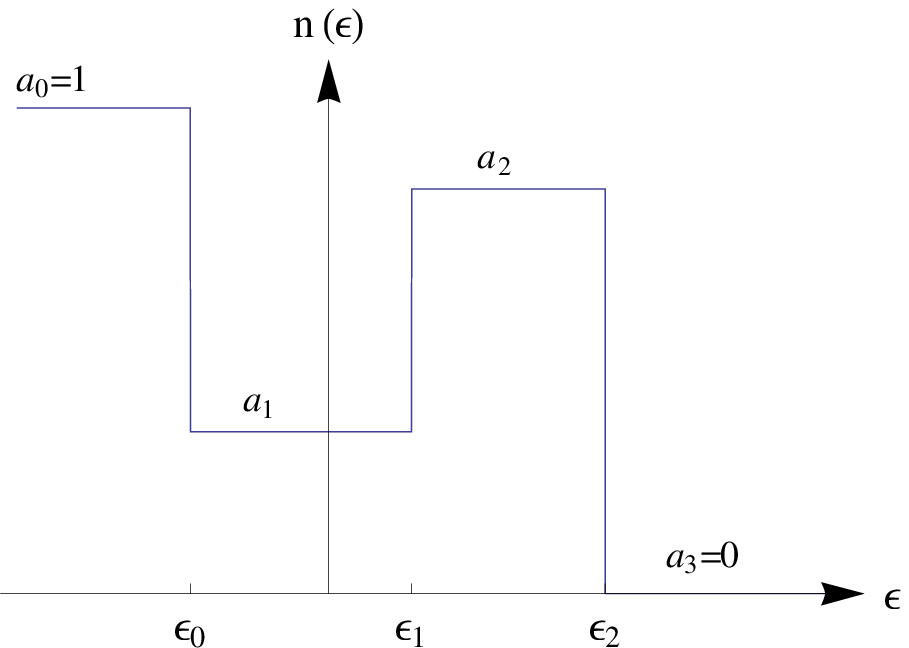}
\includegraphics[width=190pt]{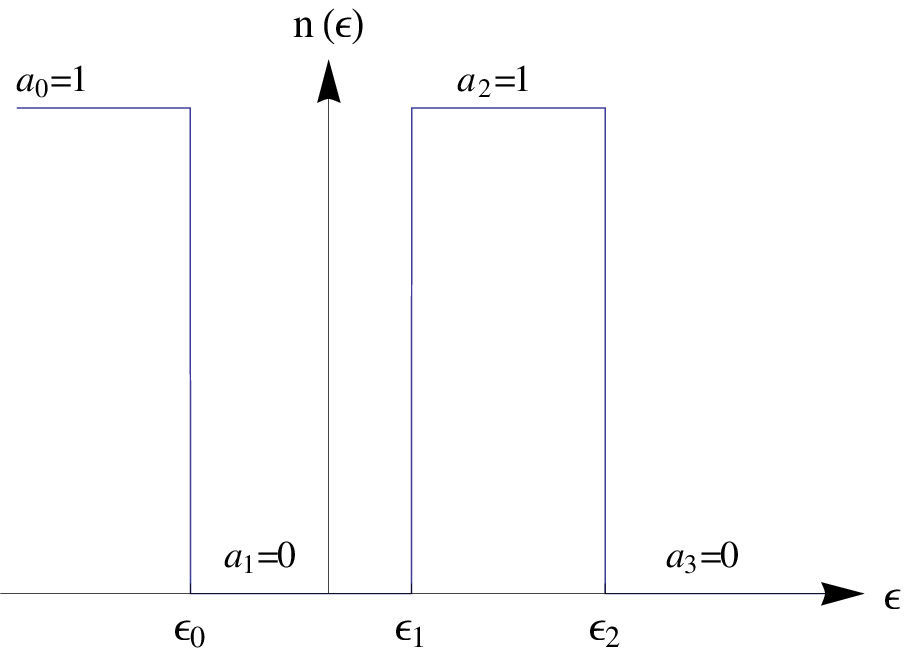}
\caption{\small {\it Left panel.} Triple-step distribution: an example
  of multi-step distribution (\ref{multistep-distribution}). {\it
    Right panel.}  Simplest non-trivial distribution of the type
  (\ref{multistep-distribution}) with identically
  zero dephasing.} 
\label{Fig:multistep-distribution}
\end{figure*}

\vspace*{1.4cm}

Let us now consider a distribution function with multiple steps:
(cf. Fig. \ref{Fig:multistep-distribution}):  
\begin{equation}
\label{multistep-distribution}
n(\epsilon) = \left\{
\begin{array}{ll}
1 \equiv a_0 \,, & \qquad \epsilon < \epsilon_0 \\
a_1\,, & \qquad \epsilon_0 < \epsilon <\epsilon_1 \\
\ldots &  \\
 a_m\,, & \qquad \epsilon_{m-1} < \epsilon <\epsilon_m \\
0 \equiv a_{m+1}\,, & \qquad \epsilon_m < \epsilon\,,  
\end{array}
\right.
\end{equation}
where $0\le a_j \le 1$, for $j=0,\dots,m$, 
We are interested in the Toeplitz determinant 
({\ref{Toeplitz}}) for  the multi-step distribution function 
(\ref{multistep-distribution}). Let us split the phase $\delta$ as 
$\delta=2\pi M+\delta'$, where $M$ is integer and $|\delta'| < \pi$.   
We find it convenient to normalize the determinat by  its zero-temperature value 
(\ref{DeltN0}); the normalized determinant will be denoted as
$\overline{\Delta}_\tau(\delta)$.  
According to the extended Fisher-Hartwig conjecture \cite{Gutman10}, 
it has the following long time asymptotics

\begin{widetext}
\begin{eqnarray}
 \overline{\Delta}_\tau(\delta) &=&  \frac{\exp\left[
-i\tau \mu'-\tau/2\tau_\phi\right]}{G(1-\delta/2\pi)G(1+\delta/2\pi)}
 \nonumber \\
& \times & \sum_{n_0+\ldots+n_m=-M}
e^{i\tau \sum_j n_j \epsilon_j}
\left. \prod_{j<k}\left(\frac{1}{\tau U_{jk}}\right)^{-2\beta_j\beta_k}
 \prod_j
 G(1+\beta_j)G(1-\beta_j) \right|_{\beta_j=\beta'_j+n_j}\, (1 + \ldots).
\label{Delta_asymptotic}
\end{eqnarray}
\end{widetext}
Here we use the following notations:
the exponents $\beta_j'$ (satisfying $|{\rm Re} \beta_j'|<1/2$) are 
\begin{equation}
\label{fh8}
\beta_i' = {i\over 2\pi  }\left[\ln(1-a_{i+1}+a_{i+1}e^{-i\delta})
-\ln(1-a_i+a_ie^{-i\delta})
\right], 
\end{equation} 
the dephasing rate  reads
\begin{equation}
\label{fh9}
{1\over \tau_\phi} =2{\rm Im}\sum_j \beta_j'\epsilon_j\,,
\end{equation} 
$G(x)$ is the Barnes G-function,  $U_{jk} = |\epsilon_j-\epsilon_k|$,  
and $\mu' = - {\rm Re} \sum_j\beta'_j\epsilon_j$. 
Note that the ultraviolet regularization $\Lambda$ does not enter the
normalized determinant.  
The asymptotic (\ref{Delta_asymptotic}) is valid provided that $\tau
U_{jk}\gg 1 $ for all $j \ne k$. The summation goes
over all sets of integer $n_0,\ldots, n_m$ satisfying $n_0+\ldots +n_m
= -M$; each such set yields the corresponding oscillatory exponent
$e^{i\tau \sum_j n_j \epsilon_j}$. Equation (\ref{Delta_asymptotic})
presents explicitly the leading asymptotic behavior for the factor
multiplying each of these exponents. Apart from this dominant term,
there will be in general also subleading (in powers of $1/t$) terms
corresponding to the same exponent; these are abbreviated by $+\ldots$
in the last bracket.  

The asymptotics (\ref{Delta_asymptotic}) has  a long history. The form of
its leading term (the one with the slowest decay in $\tau$, i.e. with the
smallest   exponent  
$\alpha(n_0\,,\ldots n_m)=2\Real\left[\sum_{j>k}\beta_j\beta_k \right]$)
was suggested back in 1968 by Fisher and Hartwig
\cite{fisher-hartwig}. Since then, significant efforts were invested
into the exact formulation, the proof, and extensions of the Fisher-Hartwig
conjecture.  For the case when a unique combination of integers $n_i$
exists that minimizes the exponent $\alpha(n_0\,,\ldots n_m)$, 
the leading asymptotic term (including the corresponding numerical 
coefficient indicated in (\ref{Delta_asymptotic})) was rigorously
derived by Ehrhardt\cite{Ehrhardt}. In a recent seminal
paper \cite{deift09} the theorem due to Ehrhardt was generalized for the
case when there are several distinct  
sets of integers $\{n_i\}$ sharing the same minimal value of the
exponent $\alpha(n_0\,,\ldots n_m)$.  
It was proven that the {\it leading} term of the asymptotic expansion of the determinant 
$\Delta_{\tau}[\delta]$ at large $\tau$ is given by
(\ref{Delta_asymptotic}) where the sum should be restricted  
to the sets $\{n_i\}$ minimizing the exponent $\alpha(n_0\,,\ldots n_m)$.

More recently, two of us and Gefen \cite{Gutman10} formulated and extended version of
the Fisher-Hartwig conjecture [as shown in Eq. (\ref{Delta_asymptotic})] that
includes a sum over all sets $\{n_i\}$ (which correspond to different
branches of the logarithms) and
captures the leading term of the expansion {\it at every} oscillation frequency 
$\sum_j n_j\epsilon_j$. This extension is very natural from
the point of view of continutiy, as, under change of parameters, the
dominant branch (that determines the leading asymptotics given by
Ref.~\onlinecite{deift09}) may become subdominant. This is particularly
transparent in the energy representation of our problem discussed
below: different branches then correspond to singularities near
different energies; clearly, such a singularity will persist even when
its exponent will become subdominant with respect to a singularity at
other energy. Furthermore, the summation over branches has a clear
physical meaning: it corresponds to processes including transfer of
one or several electrons between different Fermi edges \cite{Gutman10}. 

To illustrate how Eq.~ (\ref{Delta_asymptotic}) works, let us consider
a simple  case of the determinant at the phase $\delta=4\pi$,  
which can be evaluated exactly by a  ``refermionization'' procedure
\cite{GGM_long2010}. We assume for definiteness  that the distribution
function is of double step form with jumps at $\epsilon_0$ and 
$\epsilon_1$. (Generalization to a distribution function with more
than two jumps is straightforward.) The exact result reads 
\begin{multline}
\label{Delta4pi}
 \overline{\Delta}_\tau[4\pi]=a_1(a_1-1)
 \left[(\epsilon_1-\epsilon_0)^2\tau^2-2\right]e^{-i(\epsilon_0+\epsilon_1)\tau} 
\\+
(1-a_1)^2 e^{-2i\epsilon_0\tau}+
a_1^2
e^{-2i\epsilon_1\tau}\,.
\end{multline}
On the other hand, considering the expansion (\ref{Delta_asymptotic}) one gets
at $\delta\rightarrow 4\pi$
\begin{eqnarray}
\beta_0'=(a_1-1)\frac{\delta'}{2\pi}\\
\beta_1'=-a_1\frac{\delta'}{2\pi}
\end{eqnarray}
with $\delta'\rightarrow 0$. Observing now that $G(x)$ has $k$-th
order zero at $x=-k+1$ for any positive integer $k$
we conclude that at $\delta=4\pi$ (or generally for any $\delta$ being
integer multiple of $2\pi$) the sum in (\ref{Delta_asymptotic})
becomes finite. In the present case only the terms with  
$(n_0\,,\, n_1)=(-1, -1)$, $(-2, 0)$ and $(0, -2)$ contribute,
yielding  
\begin{multline}
\label{Delta4piA}
 \overline{\Delta}_\tau(4\pi)
 \simeq a_1(a_1-1) (\epsilon_1-\epsilon_0)^2\tau^2 e^{-i(\epsilon_0+\epsilon_1)\tau}
\\+
(1-a_1)^2 e^{-2i\epsilon_0\tau}+
a_1^2
e^{-2i\epsilon_1\tau}\,.
\end{multline}
Comparing this asymptotic formula to the exact result (\ref{Delta4pi}), we see that
Eq.~(\ref{Delta4piA}) indeed perfectly reproduce leading factors for
each oscillation frequency. The only term missing in Eq.~(\ref{Delta4piA})
is 
\begin{equation}
\label{e3.1}
-2 a_1(a_1-1) e^{-i(\epsilon_0+\epsilon_1)\tau} \,,
\end{equation}
which represents a small correction (due to an additional factor $\propto \tau^{-2}$)
to the leading term at the same frequency $\epsilon_0+\epsilon_1$,
\begin{equation}
\label{e3.2}
 a_1(a_1-1) (\epsilon_1-\epsilon_0)^2\tau^2
 e^{-i(\epsilon_0+\epsilon_1)\tau} \,.
\end{equation}
Such terms representing small power-law corrections to the leading
contribution at the same frequency are indicated in
Eq.~(\ref{Delta_asymptotic}) by the symbol  $+\ldots$. 

Let us note that, while being small with respect to the leading term
at the same frequency, these corrections are not necessarily small
with respect to leading terms at other frequencies. 
In particular, in the considered example the correction term
on the frequency $\epsilon_0+\epsilon_1$, Eq.~(\ref{e3.1}) is of the
same order  as the terms oscillating with  frequencies $2\epsilon_0$ and $2\epsilon_1$ 
that are taken into account by Eq.~ (\ref{Delta4piA}).

Thus, Eq.~(\ref{Delta_asymptotic}) captures explicitly the leading 
term for each frequency.  A mathematically rigorous proof of this
generalized form of the Fisher-Hartwig conjecture remains to be developed.
Also, one may hope that 
it is possible to generalize  Eq.(\ref{Delta_asymptotic}) 
further,  accounting also for sub-leading contribution (indicated as
$+\ldots$ in Eq.~(\ref{Delta_asymptotic})).
A construction of such a full asymptotic expansion of the Toeplitz determinant   
was discussed very recently in  Ref.~\onlinecite{Ivanov2011} for 
the special case  
$f(\epsilon)=1+(e^{-i\delta}-1)\Theta\left(U-|\epsilon|\right)$, where
$\Theta(x)$ is the Heaviside theta function. 

It is worth mentioning that  for $\delta=2\pi$ 
Eq.~(\ref{Delta_asymptotic}) reproduces the exact result
\begin{equation}
\label{e3.3}
 \overline{\Delta}_{\tau}(2\pi)=(1-a)e^{-i\epsilon_0\tau}+ae^{-i\epsilon_1\tau}
\end{equation}
without any corrections at all. While Eq.~(\ref{e3.3}) is written for
a double-step distribution, this statement is valid for any multi-step
distribution as well. The only non-zero terms in
Eq.~(\ref{Delta_asymptotic}) for $\delta=2\pi$ are those with all
$n_j$ being equal to zero except for one equal to $-1$. The 
determinant demonstrates oscillations at frequencies $\epsilon_j$. All
corrections of the type $+\ldots$ in Eq.~(\ref{Delta_asymptotic})
vanish. This implies that for  values of $\delta \simeq 2\pi$  
the correction terms $+\ldots$ in Eq.~(\ref{Delta_asymptotic})  
have additional smallness.

Having clarified the status of the expansion (\ref{Delta_asymptotic}), let us 
now discuss its implications. In a  generic case, the power-law decay of
$\overline{\Delta}_\tau(\delta)$ is cut off by the non-equilibrium
dephasing time $\tau_\phi$ given by Eq.~(\ref{fh9}). Quite remarkably, the
dephasing time is an oscillating function of the phase $\delta$ which
translates, e.g., into the non-monotonous dependence of $\tau_\phi$ on
the interaction strength in Luttinger liquid \cite{GGM_long2010}. The
dephasing is absent when $\delta'=0$ in which case
$\Delta_{\tau}(\delta)$ can be represented in terms of a free fermionic theory.   

Dephasing is also absent for the case when all $a_j=0,\,1$.
This corresponds to the case of a {\it pure} electronic state
(i.e. characterized by a wave function rather than by a density matrix). 
The simplest non-trivial distribution
of the type Eq.~(\ref{multistep-distribution})
that has this property is the triple-step distribution of
Fig. \ref{Fig:multistep-distribution} (right panel). We stress that in
this ``ideal inverse population''  
case the dephasing rate is  identically equal to zero, regardless 
of the value of the  phase $\delta$. Apart from being interesting on a pure 
theoretical grounds, the distributions realizing the inverse population
of electronic states are  also expected to be experimentally relevant,  
as they are inevitably generated in course of
evolution of a smooth perturbation of electronic density
if the spectral curvature is taken into account \cite{Wiegmann}.

The power-law decay of $\overline{\Delta}_\tau(\delta)$ in the time domain
is translated into the singular energy dependence of correlation
functions in energy representation. Specifically, every term in the
expansion  
(\ref{Delta_asymptotic}) gives rise to a singular contribution
\begin{widetext}
 \begin{equation}
 \Real\left[e^{-i\frac{\pi}{2}(\gamma+1)}
\frac{\prod_j
  G(1+\beta_j)G(1-\beta_j)}{G(1-\delta/2\pi)G(1+\delta/2\pi)}\Gamma\left(\gamma_{n_0\ldots
  n_m}\right) 
\prod_{j<k}\left(\frac{1}{U_{jk}}\right)^{-2\beta_j\beta_k}
 \frac{1}{\left(i\epsilon+i\sum_j
   n_j\epsilon_j-i\mu'-\tau/2\tau_\phi\right)^{\gamma_{n_0\ldots n_m}}} 
\right] 
\label{singularity_Fourier}
 \end{equation}
\end{widetext}
 to the Fourier transform of
 $\overline{\Delta}_\tau[\delta]/(i\tau)^{\gamma+1}$ with the exponent  
\begin{equation}
 \gamma_{n_0\ldots n_m}=\gamma-\left.2\sum_{j<k}\beta_j\beta_k\right|_{\beta_j=\beta_j'+n_j}\,.
\label{exponet}
\end{equation}
In Sec.~\ref{Sec:NumericalAnalysis} we will compare Eqs.~(\ref{singularity_Fourier}),
(\ref{exponet}) with the results of numerical evaluation of Toeplitz determinants.

\section{Numerical analysis}
\label{Sec:NumericalAnalysis}

In this Section, we present results of the numerical analysis of the
Toeplitz determinants (\ref{Toeplitz}) which allows us to evaluate the
many-body Green functions in the whole range of times (energies). We
will further demonstrate that the numerics gives full support to the
asymptotic expansion (\ref{singularity_Fourier}), (\ref{exponet}).  

\subsection{Numerical procedure}

To be specific, we will consider fermions with the following two types
of many-step distributions: (i) double-step distribution 
\begin{equation}
\label{nd}
 n_d(\epsilon) = \left\{
\begin{array}{ll}
1 \equiv a_0 \,, & \qquad \epsilon < \epsilon_0=-U/3  \\
a_1=1/3\,, & \qquad \epsilon_0 < \epsilon <\epsilon_1=2U/3 \\
0 \equiv a_{2}\,, & \qquad \epsilon_1 < \epsilon\,,
\end{array}
\right.
\end{equation}
and (ii) triple-step distribution with the ``maximal'' inverse population
(Fig. \ref{Fig:multistep-distribution}, right panel) 
\begin{equation}
\label{nt}
n_t(\epsilon) = \left\{
\begin{array}{ll}
1 \equiv a_0 \,, & \qquad \epsilon < \epsilon_0=-3U/4 \\
a_1=0\,, & \qquad \epsilon_0 < \epsilon <\epsilon_1=-U/2 \\
a_2=1\,, & \qquad \epsilon_1 < \epsilon <\epsilon_2=U/4 \\
 0 \equiv a_{3}\,, & \qquad \epsilon_2 < \epsilon\,.  
\end{array}
\right.
\end{equation}
In these equations we have expressed all the energies $\epsilon_k$
in terms of characteristic scale $U=\epsilon_m-\epsilon_0$ associated
with the distribution function.  

\begin{figure*}
 \includegraphics[width=500pt]{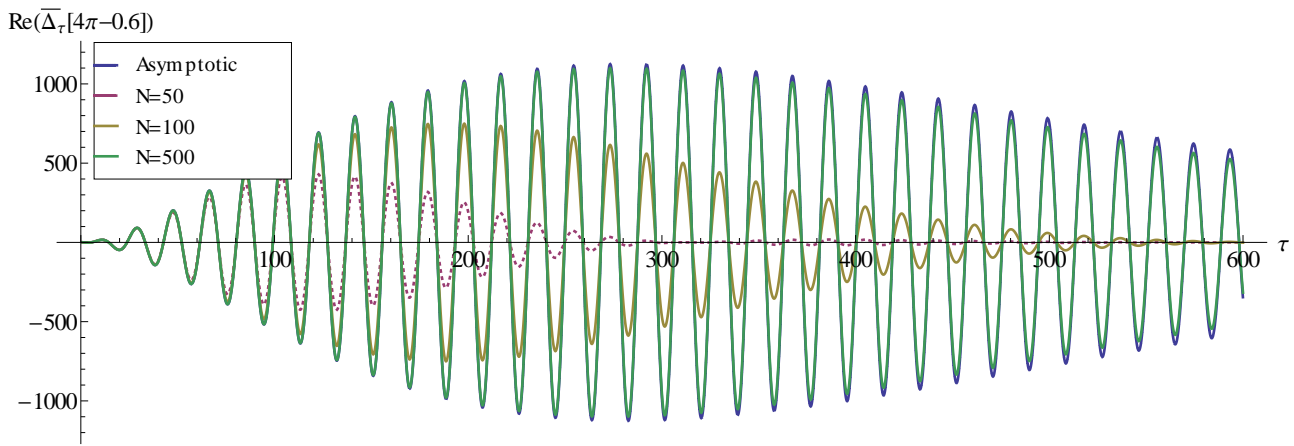}
\caption{\small Time dependence of the determinant
  $\overline{\Delta}_\tau[4\pi-0.6]$. The distribution function  
was taken to be $n_d(\epsilon)$, Eq. (\ref{nd}). Numerical results for
matrix sizes $N=50$, 100, and 500, as well the leading asymptotic term
(\ref{DeltaA1}) are shown. The
numerical result for $N=500$ is almost indistinguishable from the
asymptotics.   
 }
\label{Fig:TimeDomain}
\end{figure*}

Let us consider the normalized determinant
$\overline{\Delta}_\tau[U,\delta]$ and its finite-dimensional
approximation 
$\overline{\Delta}_N[U/\Lambda, \delta]$. Here we made explicit the
dependence of the determinants on $U$. 
At $\tau$ and $U$ fixed, $\overline{\Delta}_N[U/\Lambda, \delta]$ has
a finite limit as $\Lambda\rightarrow \infty$ 
which is a cutoff-independent function of the dimensionless variable  
$U\tau$ only: 
\begin{eqnarray}
 \overline{\Delta}_\tau[U, \delta]\equiv\overline{\Delta}[U\tau,
   \delta]
&=& \lim_{\Lambda\rightarrow\infty}
\overline{\Delta}_{N=\frac{\tau\Lambda}{\pi}}\left[\frac{U\tau}{\pi N}, \delta\right]\\
&=&
\lim_{N\rightarrow\infty}
\overline{\Delta}_{N}\left[\frac{U\tau}{\pi N}, \delta\right]\,.
\label{DoubleScalingLimit}
\end{eqnarray}

Equation (\ref{DoubleScalingLimit}) constitutes the starting point for
our numerical analysis. With a simple  
{\it Mathematica} code we are able to go within a quite short
computation time up to the size of the
Toeplitz matrix $N=500$, which is typically sufficient for the convergence  
to the large-$N$ limit for relevant values of $U\tau$. 

The convergence properties of our procedure become generally worse at
large $\delta$. Thus, we chose to illustrate them  
with the calculation of the determinant at the phase $4\pi-0.6$ which is
larger then any  phase we will encounter in the next section.  This
choice also enables us to demonstrate clearly the presence of
the correction terms indicated by $+\ldots$ in  Eq.~(\ref{Delta_asymptotic}). 
    
From now on,  we measure $\tau$ in units of $1/U$.
Figure \ref{Fig:TimeDomain} shows the  result of numerical evaluation
of the normalized Toeplitz determinant
$\overline{\Delta}_{\tau}[4\pi-0.6]$   
for  the double-step distribution function $n_d$ given by Eq. (\ref{nd}). 
We have plotted the data for $N=50, 100,  500$ together with
leading term of the asymptotic expansion~(\ref{Delta_asymptotic}), the
one with $n_0= n_1=-1$ 
\begin{equation}
 \overline{\Delta}_\tau^{A1}[4\pi-0.6]\approx (0.25 + 0.026i)
 e^{-i\tau(\epsilon_1+\epsilon_0)-\tau/2\tau_\phi} 
\tau^{1.81}\,.
\label{DeltaA1}
\end{equation}
Here $\tau_\phi\approx 77$ and $\epsilon_0+\epsilon_1=1/3$.
Note the fast convergence with the increase of the matrix size
and perfect agreement with  the predicted asymptotic behavior.  We
stress that the asymptotic fit used here has no adjustable 
parameters.  

\begin{figure}
 \includegraphics[width=210pt]{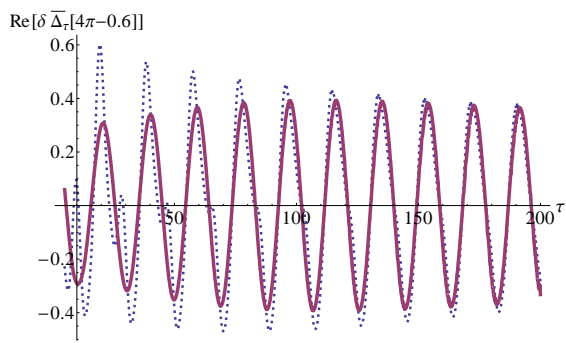}
\caption{\small Difference between the numerically evaluated  determinant 
(shown in Fig.~\ref{Fig:TimeDomain}) 
and the asymptotic approximation for $N=5000$.  
The electronic distribution was the
double-step distribution $n_d(\epsilon)$.  
Dotted line: only the leading term (\ref{DeltaA1}) was subtracted; 
full line: three main harmonics (Fisher-Hartwig branches) of the
expansion (\ref{Delta_asymptotic}) 
have been taken into account.  The remaining difference is due to a correction
[of the type indicated by $+\ldots$ in Eq. (\ref{Delta_asymptotic})]
to the leading harmonic (\ref{DeltaA1}).} 
\label{Fig:TimeDomain2}
\end{figure}

Let us now explore the effect of the other terms in the expansion
(\ref{Delta_asymptotic}). The next two terms are   
characterized by $(n_0, n_1)=(-2, 0)$ and $(n_0, n_1)=(0, -2)$.  Apart
from the exponential damping at scales larger then $\tau_\phi$ they
decay as $\tau^{-0.12}$ and $\tau^{-0.25}$ respectively. Since in this
case  powers of the  leading  and the subleading harmonics are  substantially 
different, a reliable observation of the subleading ones requires
more substantial numerical efforts.   
To achieve the required accuracy, we  use larger values of the matrix size ($N=5000$). 
Note that in Sec. \ref{FES}, \ref{TDOS}, where we focus on smaller
values of the phase shift $\delta$, subleading harmonics will be much
more pronounced and easily seen. 

The difference between the numerically calculated Toeplitz determinant (\ref{Toeplitz}) 
and its asymptotic approximation \ref{Delta_asymptotic}) is shown in
Fig. \ref{Fig:TimeDomain2}.   
The dotted line corresponds to the difference between 
the numerical result and the leading term (\ref{DeltaA1}). 
The solid  line is the difference between the numerical result and the first three terms 
in the expansion (\ref{Delta_asymptotic}).
As expected, inclusion of the terms $(n_0, n_1)=(-2, 0)$ and $(n_0,
n_1)=(0, -2)$ improves the agreement between the asymptotics and the
exact results. 
Indeed, the oscillations at high frequencies, that are clearly seen on 
a dotted line,  are absent  on  the solid line. 
Nevertheless, a clear difference between the exact result and the asymptotic 
formula remains, which is predominantly due to the correction [of the type
$+\ldots$ in Eq.~(\ref{Delta_asymptotic})]  
to the leading  (oscillating with frequency $\epsilon_1+\epsilon_2=1/3$) harmonic.

\label{FES}
\begin{figure*}
\includegraphics[width=210pt]{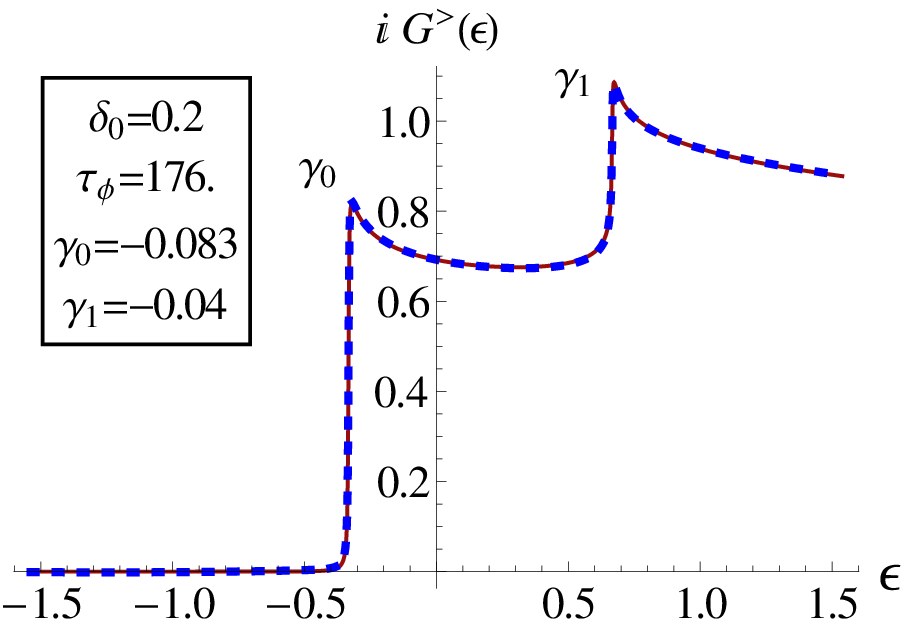}
\includegraphics[width=210pt]{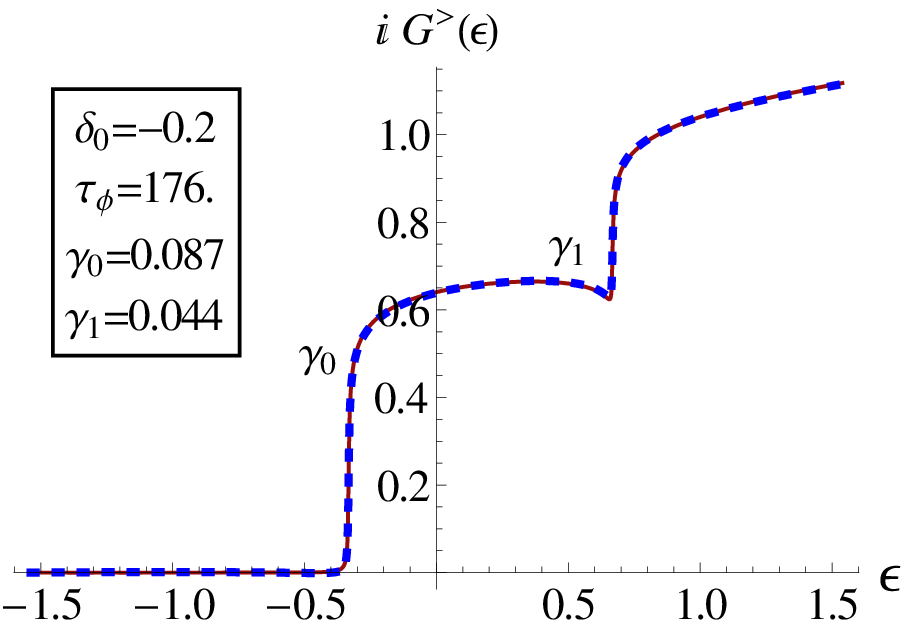}
\includegraphics[width=210pt]{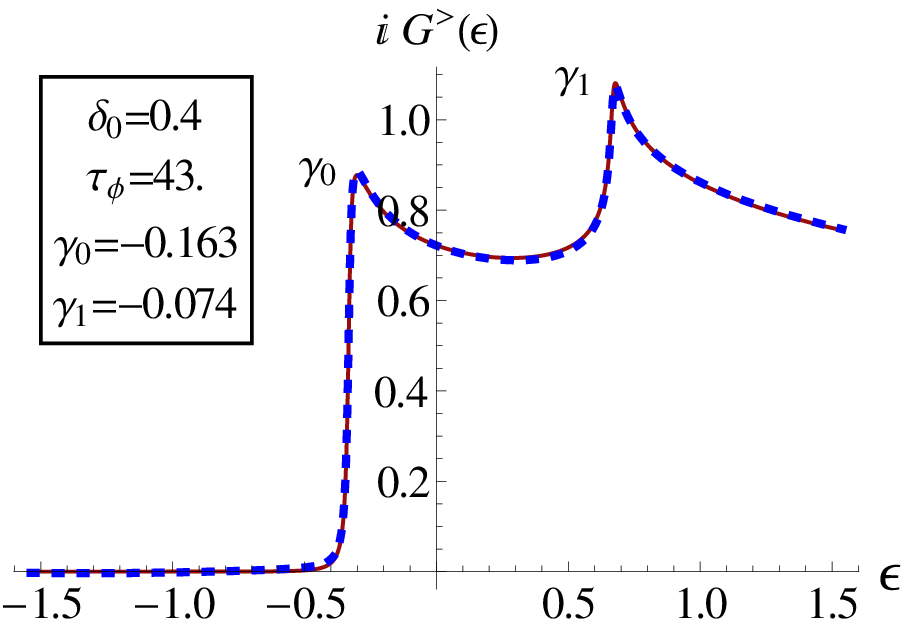}
\includegraphics[width=210pt]{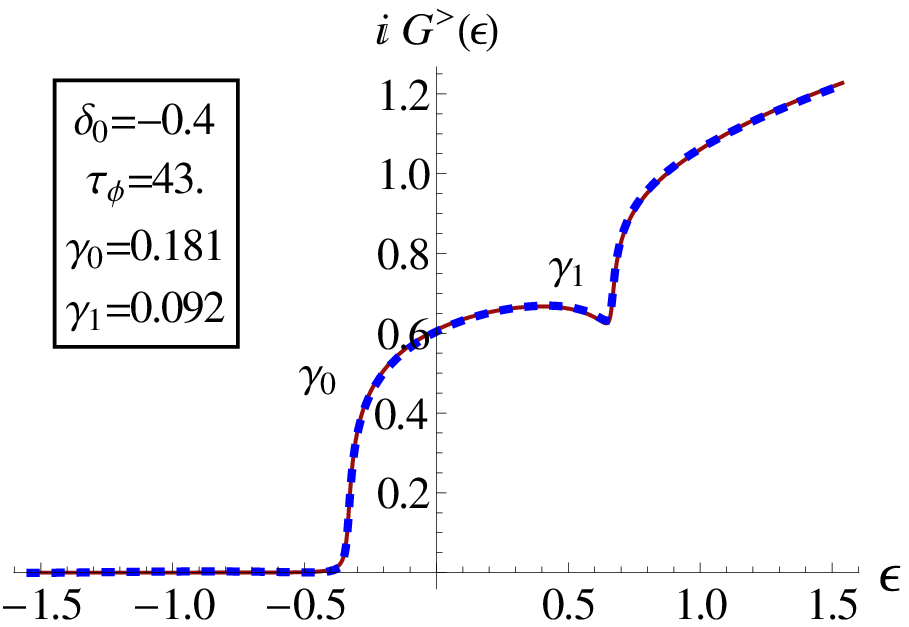}
\includegraphics[width=210pt]{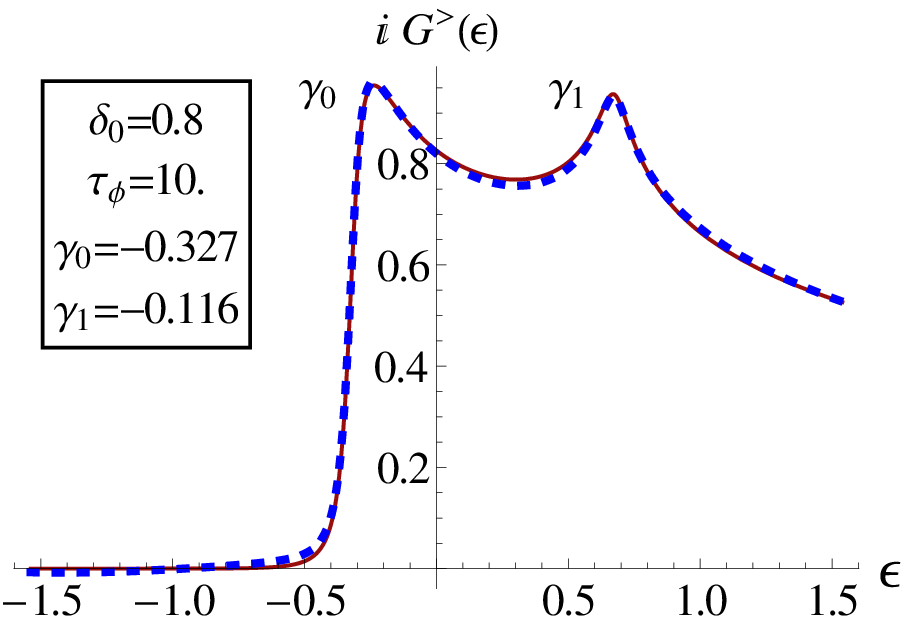}
\includegraphics[width=210pt]{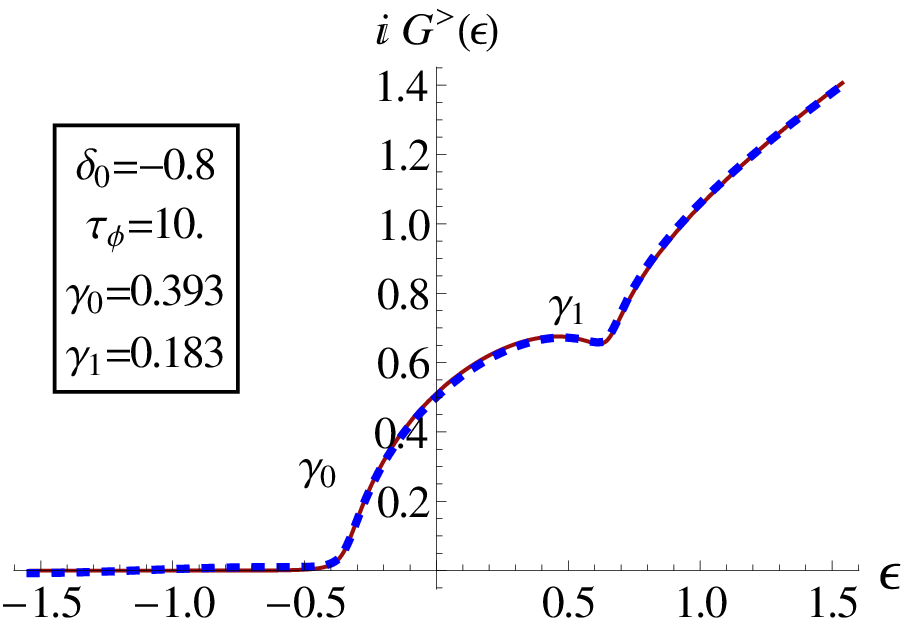}
\caption{\small X-ray absorption spectra  at different values of the
  scattering phase $\delta_0$ for the double-step  
distribution function of electrons $n_d(\epsilon)$ specified in Eq. (\ref{nd}).
 The solid lines represent the result of numerical evaluation of
 Toeplitz determinants while the dotted lines show the fits based on the
 asymptotics (\ref{Delta_asymptotic}, \ref{singularity_Fourier}). The
 legend shows the corresponding dephasing  
time  $\tau_\phi$ together with the exponents $\gamma_0$ and
$\gamma_1$ governing the singular behavior of  
$G^>(\epsilon)$ at $\epsilon=\epsilon_0=-1/3$ and $\epsilon=\epsilon_1=2/3$. }
\label{Fig:FermiEdgeDoubleStep}
\end{figure*} 

We have thus demonstrated that, even for a relatively large phase
$\delta$, the numerical simulations work perfectly and that the
large-$t$ behavior is fully understood in the framework of the
asymptotic expansion.
In the sequel, we will present the results for two physical problems
of our interest (FES and Luttinger liquid) in
the energy domain. This is more natural physically (as this
corresponds to spectroscopy measurements) and  
also gives us the possibility  to separate  the contributions of
different harmonics in (\ref{Delta_asymptotic}) within the same graph. 
We note that the Green functions
$G^{\gtrless}(\tau)$ are obtained from a Toeplitz determinant (or a
product of two Toeplitz determinants) by multiplication with
$1/(\Lambda\tau)^{\gamma+1}$ (with $\gamma$ being the zero temperature
exponent, see the last column of Table \ref{table1}). Thus, 
\begin{equation}
 G^{\gtrless}(\epsilon)=
\left(\frac{U}{\Lambda}\right)^\gamma\widetilde{G}^{\gtrless}(\epsilon/U)\,, 
\end{equation}
where the functions $\widetilde{G}^{\gtrless}(\epsilon/U)$ are cutoff independent. 
From now on we omit the energy independent  factor
$\left(U/\Lambda\right)^\gamma$ from the Green functions and measure
all the energies in  units of the characteristic scale $U$.

\subsection{Fermi edge singularity}

\begin{figure*}
\includegraphics[width=210pt]{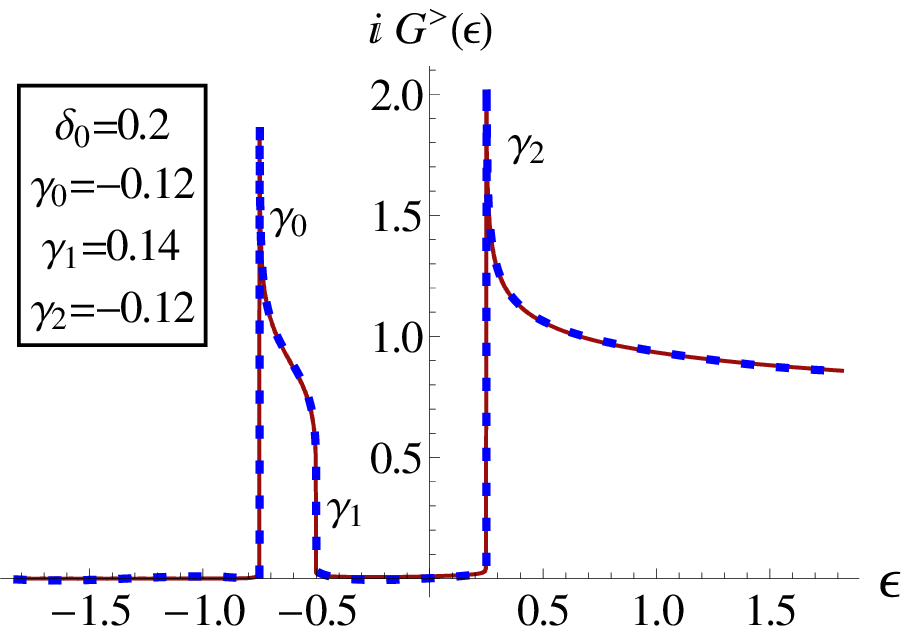}
\includegraphics[width=210pt]{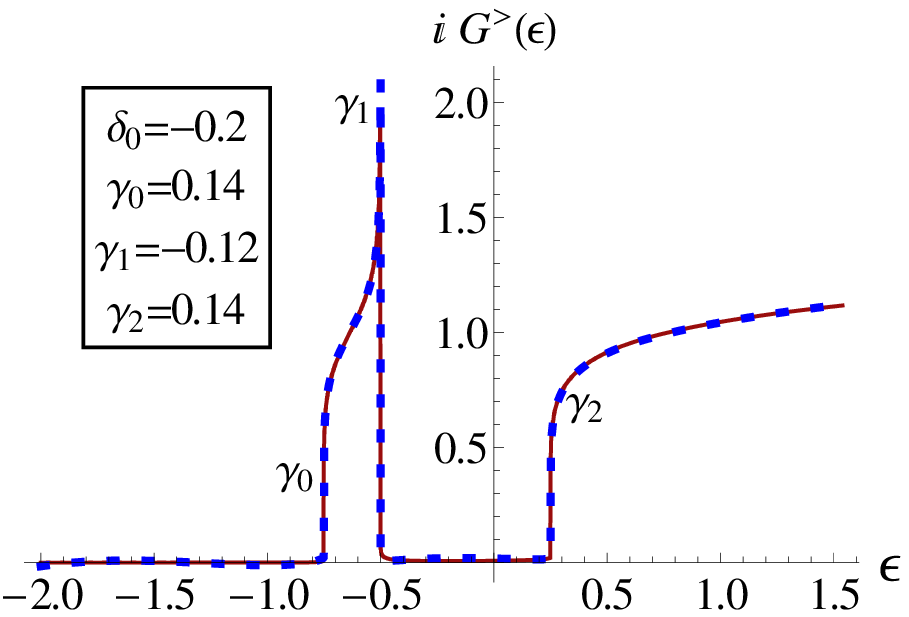}
\includegraphics[width=210pt]{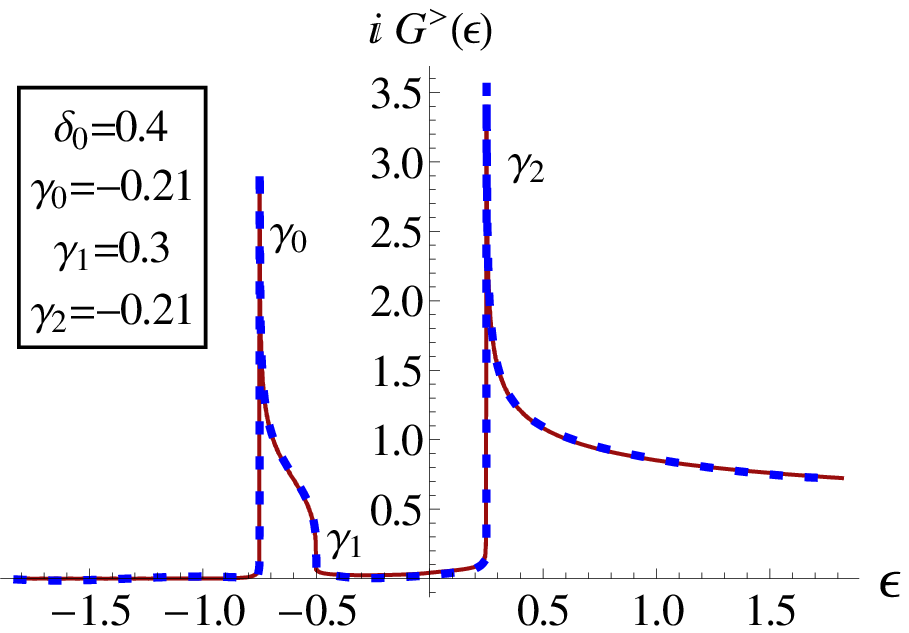}
\includegraphics[width=210pt]{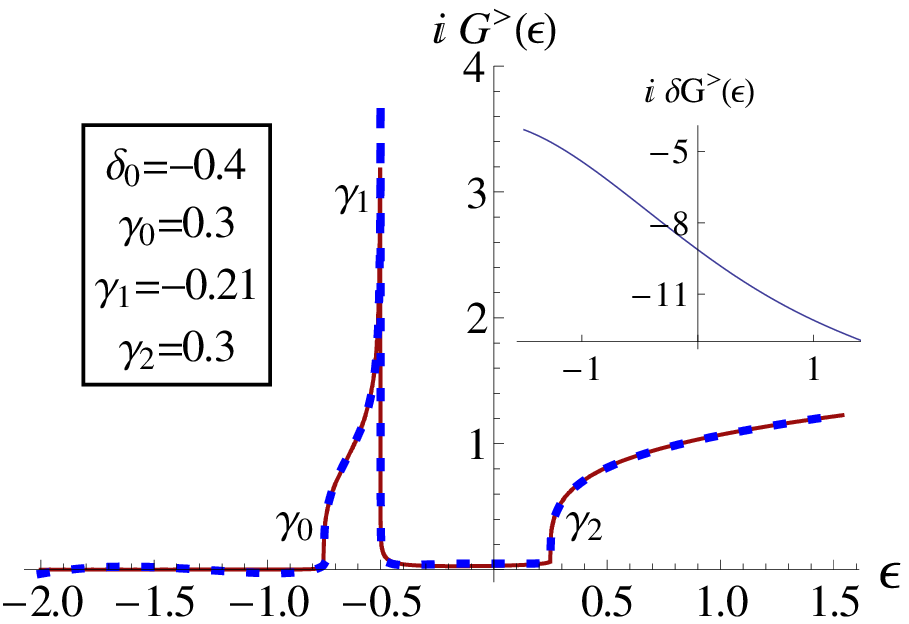}
\caption{\small X-ray absorption spectra of the non-equilibrium  FES problem
with triple-step distribution of the electrons $n_t(\epsilon)$
(see. Eq. (\ref{nt})) at relatively small scattering phase
$\delta_0 = \pm 0.2,\:\pm 0.4,\:\pm 0.8$.  
The solid lines represent the result of numerical evaluation of
Toeplitz determinants while the dots show the fits based on the
asymptotic expansion (\ref{Delta_asymptotic}).  For the chosen
distribution function the dephasing rate $1/\tau_\phi$ is identically
zero and the singularities are not smeared. The last graph in the
second column has an inset exemplifying
the smooth function $\delta G^>(\epsilon)$ added to the asymptotic
expression to fit the numerical data.
}
\label{Fig:FermiEdgeInverse}
\end{figure*}

\begin{figure*}
\includegraphics[width=210pt]{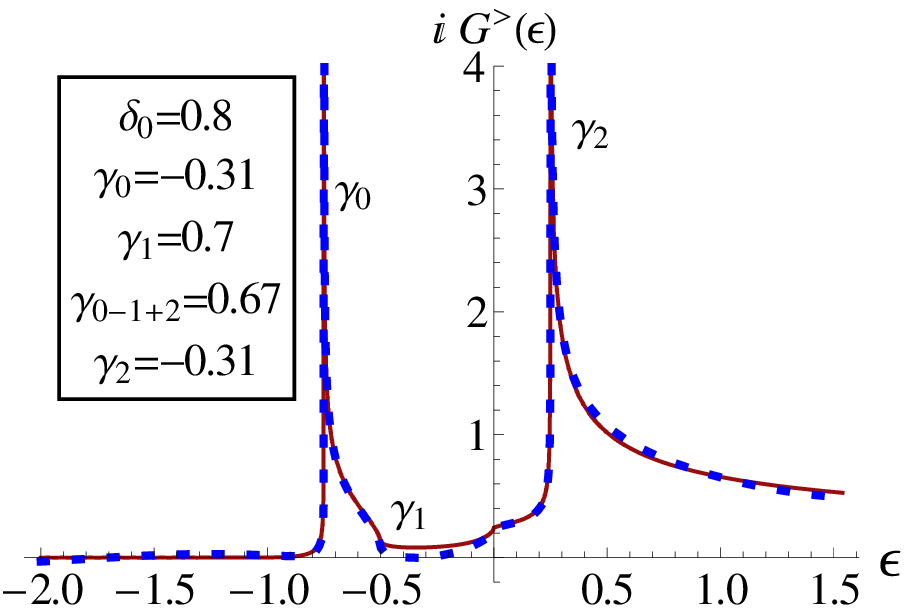}
\includegraphics[width=210pt]{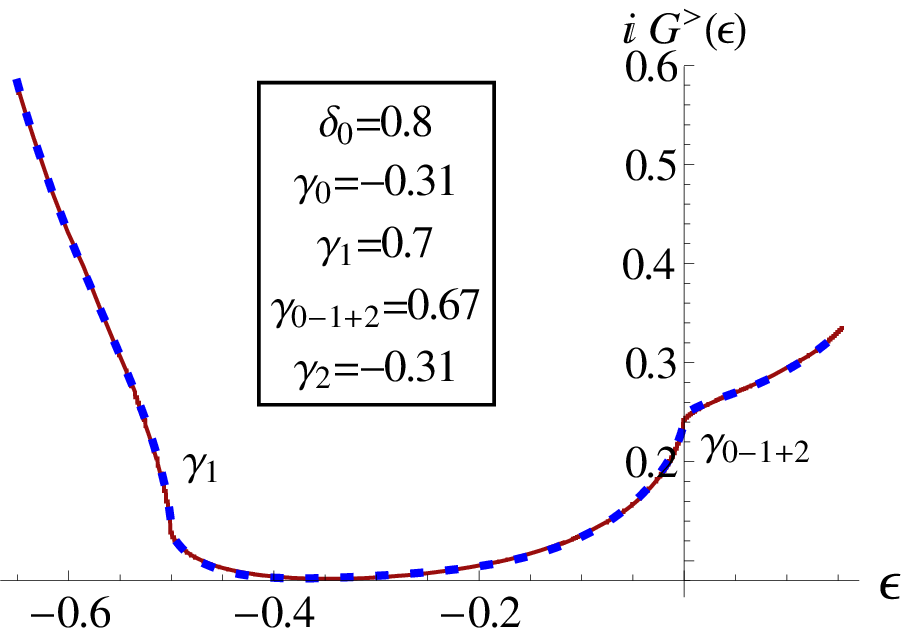}
\caption{\small X-ray absorption spectra in the non-equilibrium  FES problem
with triple-step distribution of the electrons. The electronic
distribution was the same as on Fig.  
\ref{Fig:FermiEdgeInverse} but the phase $\delta_0$ is now larger. In
addition to singularities at Fermi edge $\epsilon_k$  
one observes now a singularity at energy
$\epsilon_0-\epsilon_1+\epsilon_2=0$ with the exponent
$\gamma_{0-1+2}$ originating from the term with $n_0=-1\,,\,n_1=1\,,\,
n_2=-1$ in the sum  (\ref{Delta_asymptotic}). 
}
\label{Fig:FermiEdgeInverseAdditional}
\end{figure*}

According to Eq. (\ref{b1}), the emission/absorption rates out of
equilibrium are given by a single Toeplitz determinant.  
We analyze  the case of a double-step distribution function  
$n_d(\epsilon)$, Eq.~(\ref{nd}), first. 
The results for the different values of the scattering phase $\delta_0$
are shown in Fig.~\ref{Fig:FermiEdgeDoubleStep}. 

\begin{figure*}
\includegraphics[width=200pt]{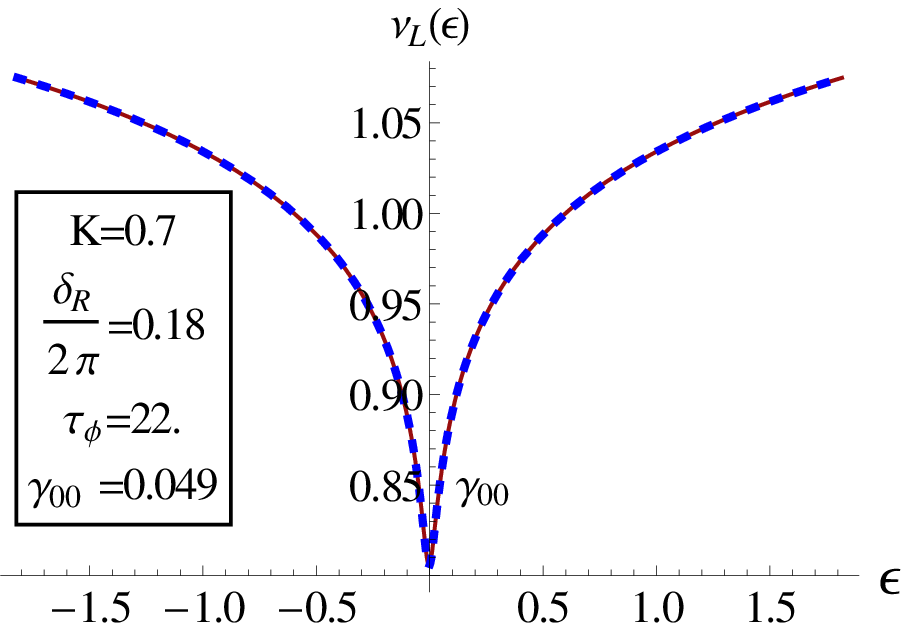}
\includegraphics[width=200pt]{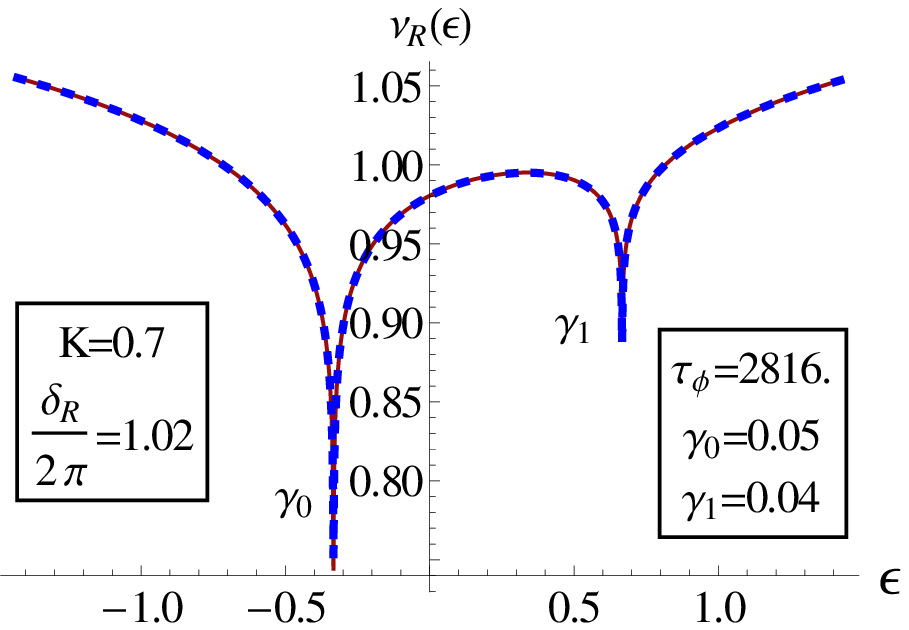}
\includegraphics[width=200pt]{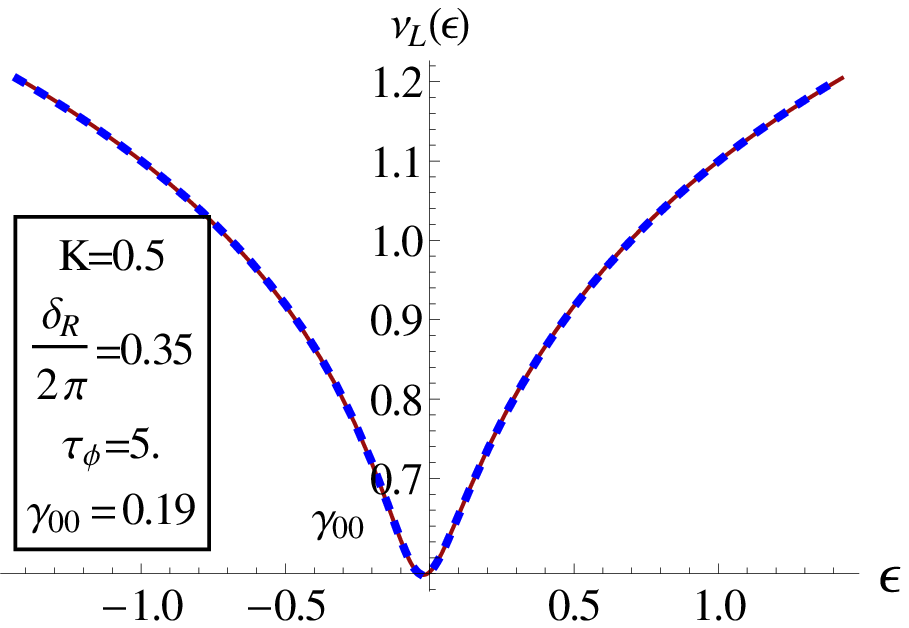}
\includegraphics[width=200pt]{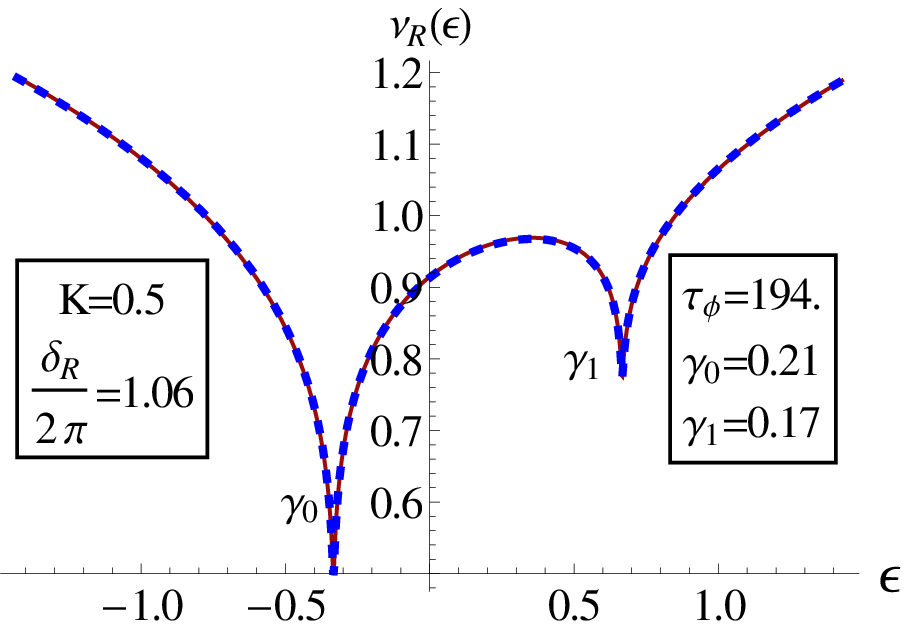}
\includegraphics[width=200pt]{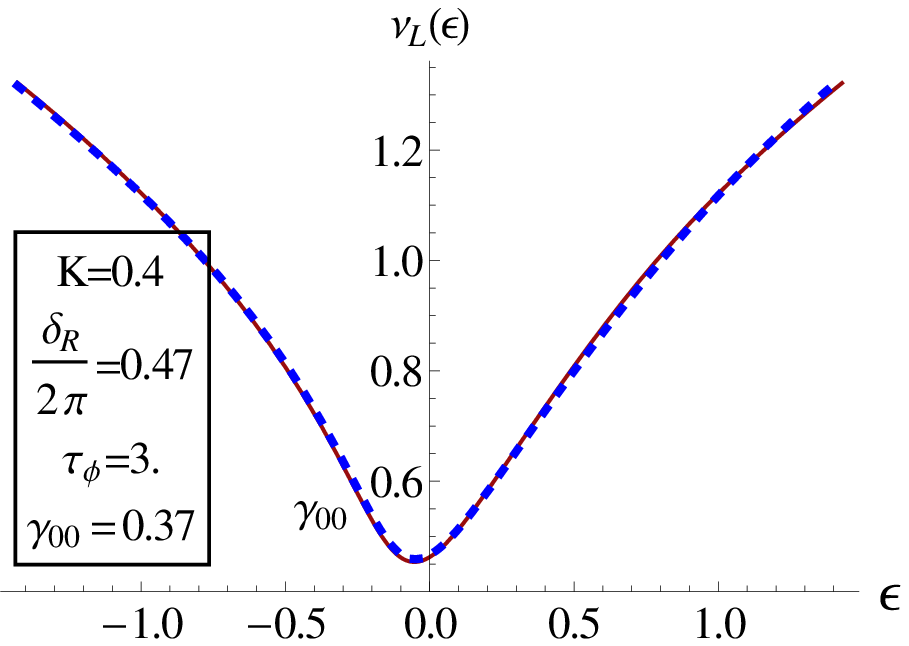}
\includegraphics[width=200pt]{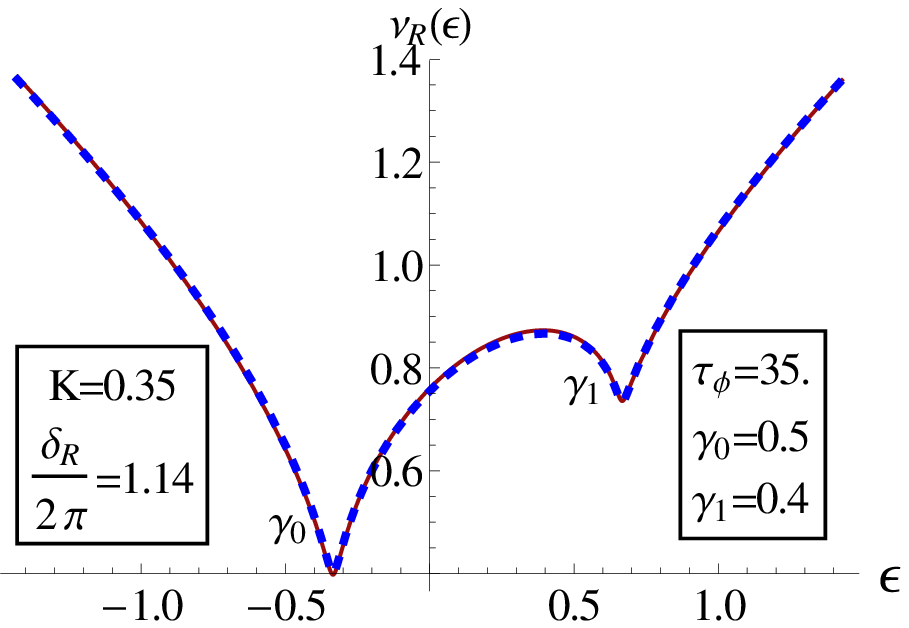}
\caption{\small TDOS $\nu_\eta(\epsilon)$ of left (first column) and
  right (second column) particles in a non-equilibrium Luttinger
  liquid. The incoming right movers have the double-step distribution
  $n_d(\epsilon)$. The incoming left movers are  assumed to have  zero
  temperature, so that $\overline{\Delta}_{L\tau}[\delta_L]\equiv 1$.
  The phase $\delta_R$ entering the nontrivial determinant 
$\overline{\Delta}_R [\delta_R]$ is in the vicnitiy of $0$ for the case of
  $\nu_L(\epsilon)$  and in the vicinity of $2\pi$ in the case of
  $\nu_R(\epsilon)$, as indicated in the legend. 
  }
\label{Fig:LuttingerLiquidDoubleStepEquilibriumLeftRight}
\end{figure*}

The solid lines in Fig.~\ref{Fig:FermiEdgeDoubleStep}
represent the result of numerical evaluation of
Toeplitz determinants, while the dotted lines show the fits based on the
asymptotic formulas (\ref{Delta_asymptotic}), (\ref{singularity_Fourier}). Only
the dominant terms in the sum  (\ref{Delta_asymptotic}) were retained
(the terms with $n_0=-1$, $n_1=0$ or vice versa). 
Using the expansion (\ref{Delta_asymptotic}), we are able to calculate  
the singular behavior of  $G^>(\epsilon)$.
The  regular part  is controlled by  the behavior of 
$G^>(\tau)$ at small $\tau$ and therefore contains the information that 
is not retained when one uses the asymptotic expressions.  
In order to compare the singular behavior predicted by the asymptotic formulas 
(\ref{Delta_asymptotic}), (\ref{singularity_Fourier})  with the 
exact  results,  we add a smooth function $\delta G^>(\epsilon)$ to the
Eqs.~(\ref{Delta_asymptotic}),
(\ref{singularity_Fourier}). We choose  $\delta G^>(\epsilon)$ in
the form of a polynomial of a relatively low order 
with coefficients that are adjusted to
optimize the fit. In fact, already a second polynomial is sufficient
to get a rather good fit, and we used it in most of the cases. In several
cases we used a fourth order poynomial.  
An example of such a  smooth function  $\delta G^>(\epsilon)$ is shown in 
Fig.~\ref{Fig:FermiEdgeInverse} (see inset of the lower right graph). 

In agreement with the analytical predictions,
the absorption spectra shown in Fig.~\ref{Fig:FermiEdgeDoubleStep}
demonstrate singular behavior $G^>_{FES}(\epsilon)\sim
\left(\epsilon-\epsilon_k+\frac{2i}{\tau_\phi}\right)^{\gamma_k}$ near
the Fermi edges $\epsilon_k\,, \,k=0, 1$. Note that the exponents at
two edges are different, which is a very good demonstration of the
importance of summation over all Fisher-Hartwig branches in  
Eqs.~(\ref{Delta_asymptotic}), (\ref{singularity_Fourier}).
One observes the enhancement of absorption
near the Fermi edges the for $\delta_0>0$.  
Contrary, for $\delta_0<0$ the absorption is suppressed. Upon increase
of the modulus of the scattering phase  
$\delta_0$, the exponents $\gamma_k$ and the inverse dephasing time
$\tau_\phi^{-1}$ grow by absolute value. Simultaneously, the dephasing
increases, which induces a stronger smearing of singularities.

In Fig.~\ref{Fig:FermiEdgeInverse}  we plot  the results 
for  triple-step distribution, $n_t(\epsilon)$, and for relatively  
small  values of the scattering phase $\delta_0$.  
At chosen $\delta_0$ the dominant terms in the expansion
(\ref{Delta_asymptotic}) are those with all $n_i=0$ except for one
$n_k=-1$ and the only visible singularities are located at the Fermi
edges $\epsilon_k$. In contrast to the case of double-step
distribution, the growth of the scattering phase $\delta_0$ is not
accompanied by smearing of the singularities, since the dephasing rate
$1/\tau_\phi$ is identically zero. 

\begin{figure*}
\includegraphics[width=210pt]{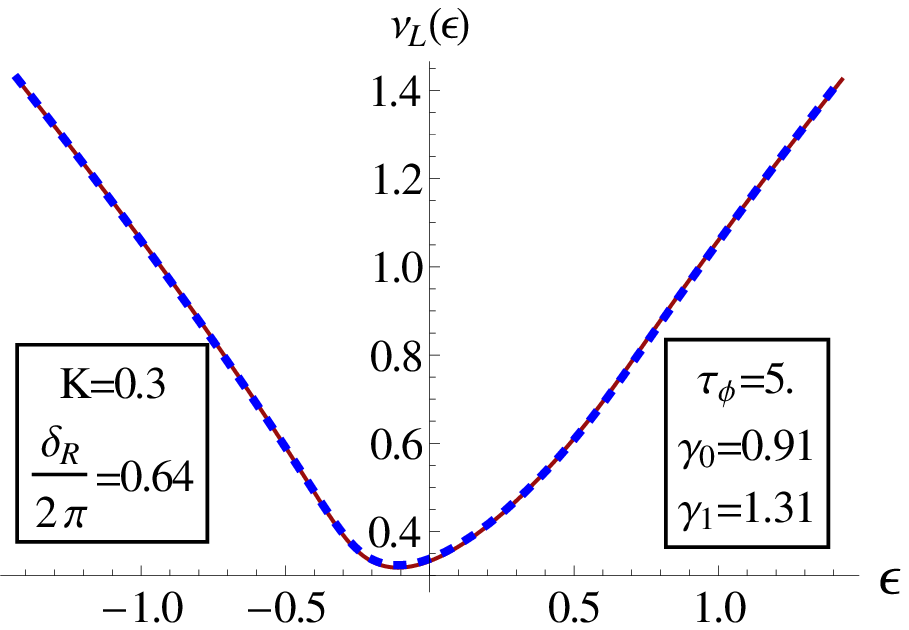}
\includegraphics[width=210pt]{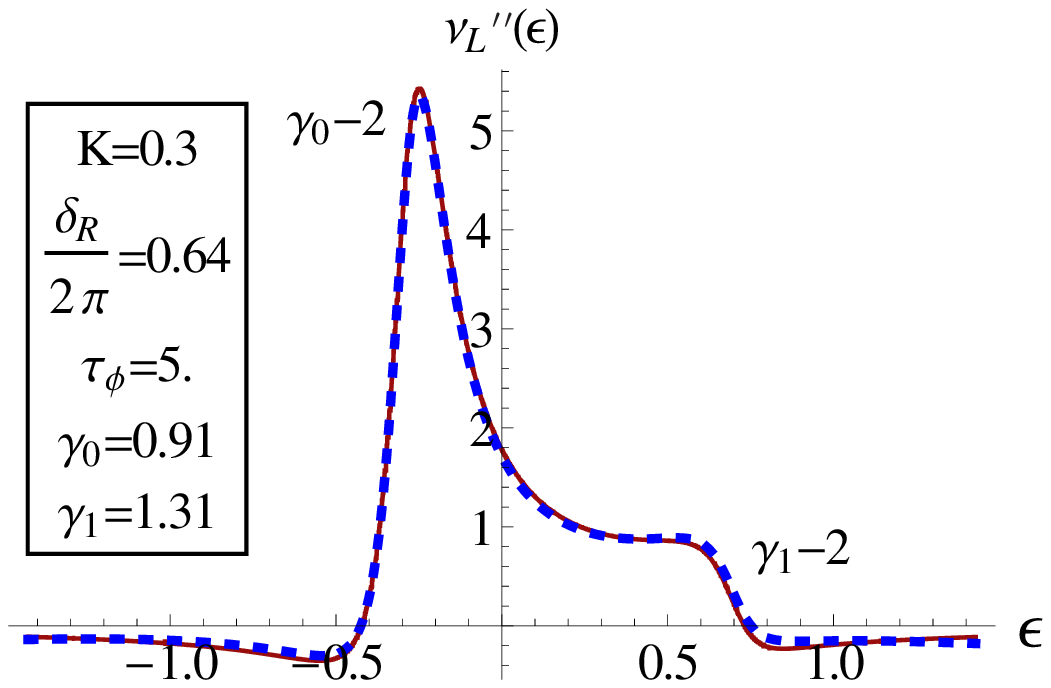}
\includegraphics[width=210pt]{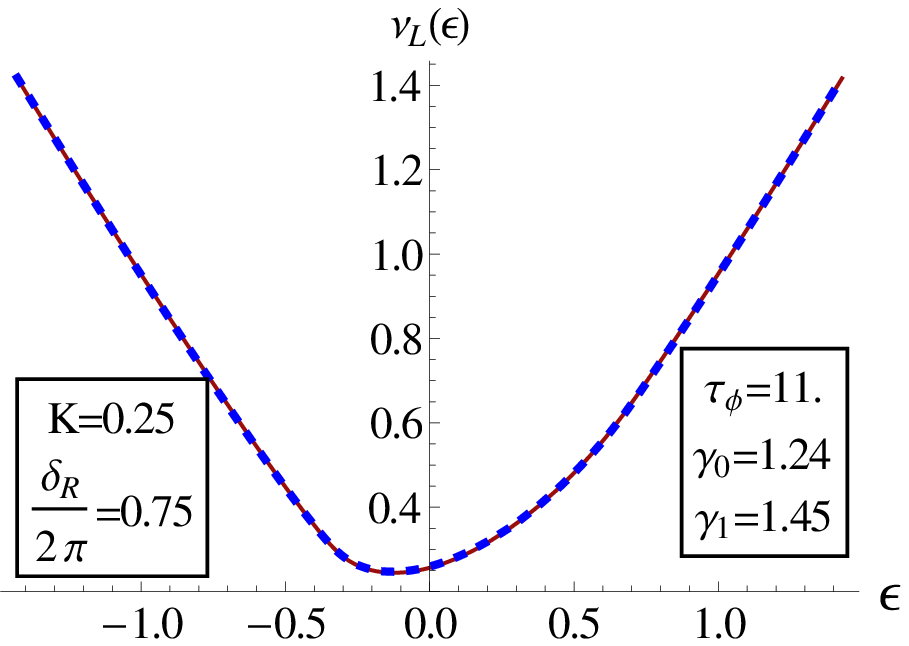}
\includegraphics[width=210pt]{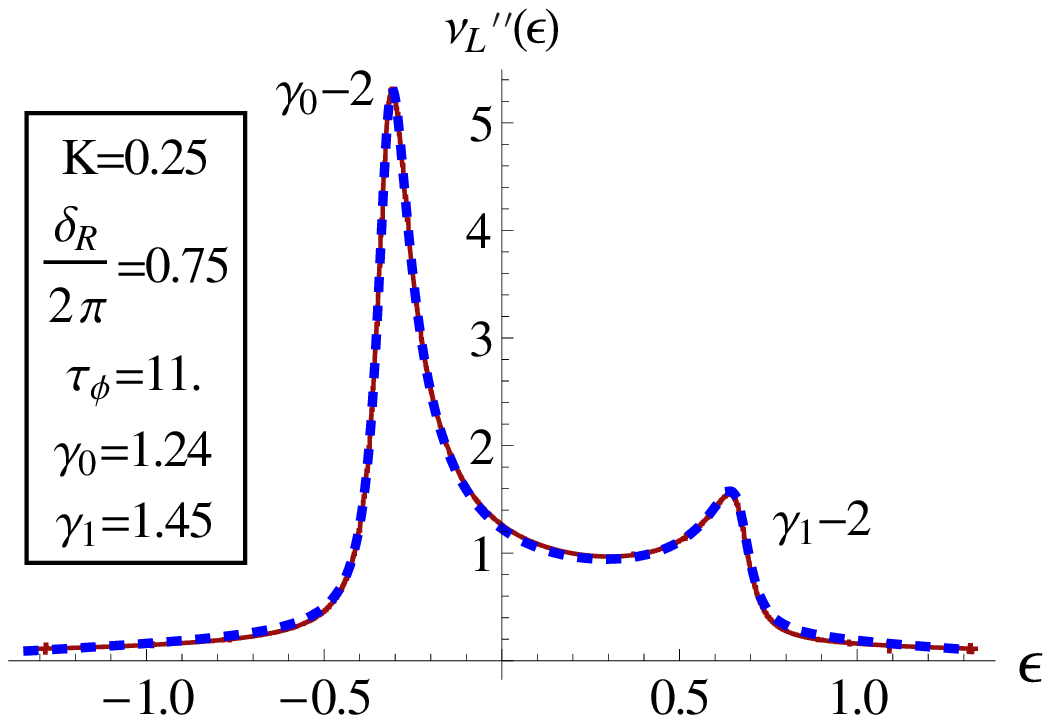}
\includegraphics[width=210pt]{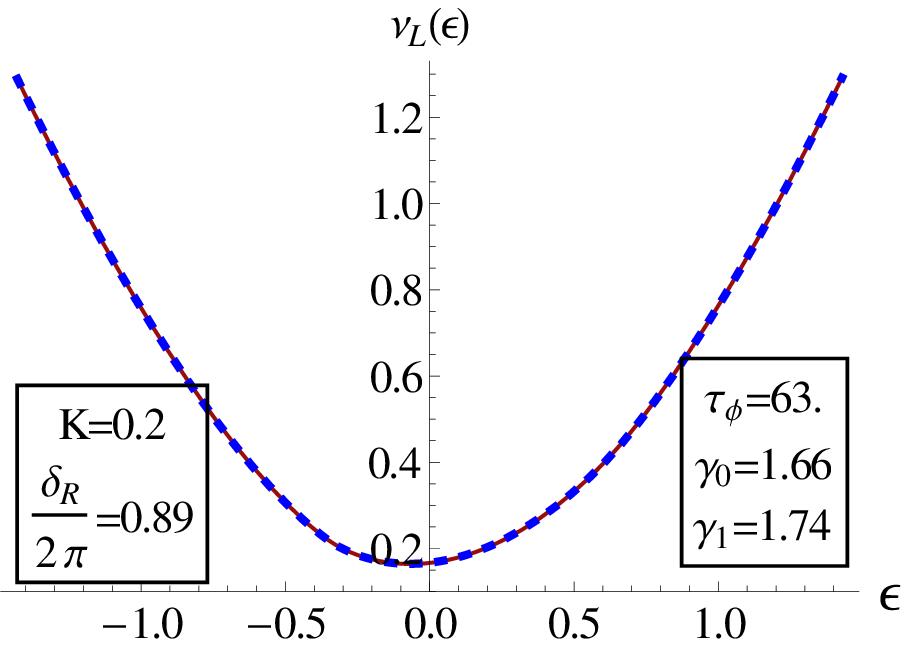}
\includegraphics[width=210pt]{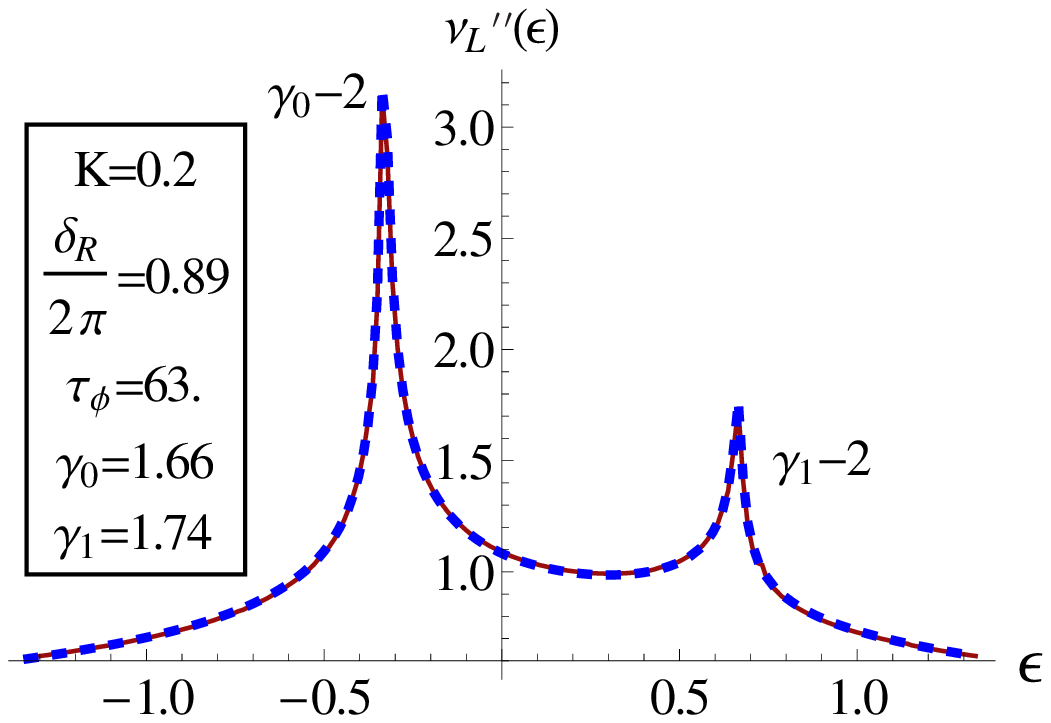}
\caption{\small Non-monotonous dependence of the non-equilibrium
  dephasing in a Luttinger liquid on the interactions strength. 
The incoming right movers have double-step distribution
$n_d(\epsilon)$. The incoming left movers are  assumed to have  zero
temperature. The interaction is now sufficiently strong so that the phase
$\delta_R$ governing the density of states for the left movers is
close to $2\pi$.  Correspondingly, the dominant singularities in
$\nu_L(\epsilon)$ are now located at $\epsilon_1$ and
$\epsilon_2$. The exponents $\gamma_0$ and $\gamma_1$ are positive and
large and the singular behavior of $\nu_L(\epsilon)$ is difficult to
see directly (left panels). It becomes evident, however,  
if one considers the second derivative of TDOS, $\nu_L''(\epsilon)$
(right panels).}
\label{Fig:LuttingerLiquidDoubleStepEquilibriumLeft}
\end{figure*}

As $\delta=2\pi-2\delta_0$ deviates further from $2\pi$, additional
terms in the series  (\ref{Delta_asymptotic})  become important, as
illustrated in Fig. \ref{Fig:FermiEdgeInverseAdditional}. In
particular, at $\delta_0=0.8$, apart from  
the singularities at Fermi edges $\epsilon_k$ a new one (at
$\epsilon=\epsilon_0-\epsilon_1+\epsilon_2=0$)  
with the exponent  $\gamma_{0-1+2}$ is clearly seen. It originates
from the term $n_0=-1\,,\, n_1=1\,,\,n_2=-1$ in
Eq.~(\ref{Delta_asymptotic}). This once more confirms the extended
Fisher-Hartwig conjecture (\ref{Delta_asymptotic}) with the summation
over all branches.

\subsection{Tunneling into non-equilibrium Luttinger liquid}
\label{TDOS}

\begin{figure*}
\includegraphics[width=210pt]{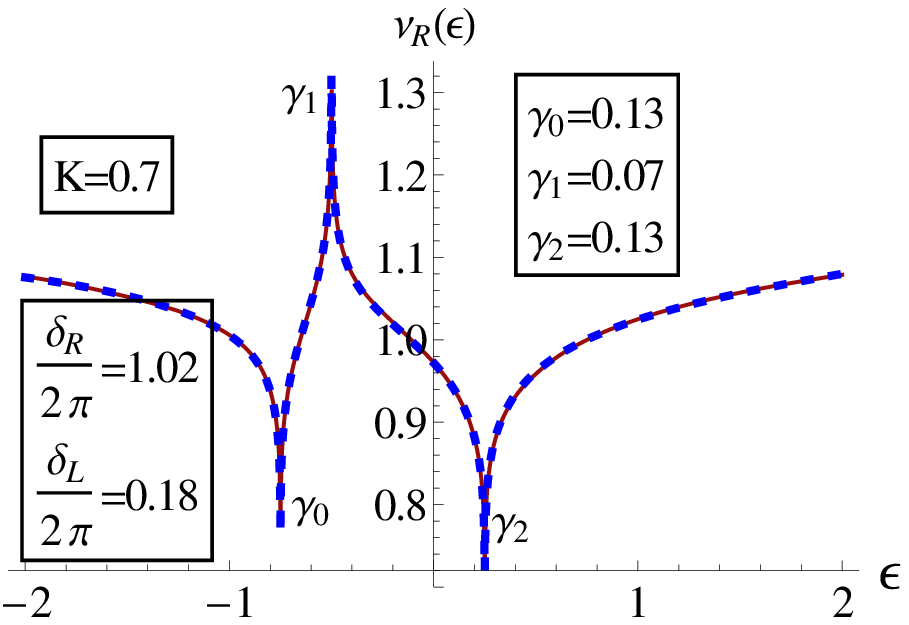}
\includegraphics[width=210pt]{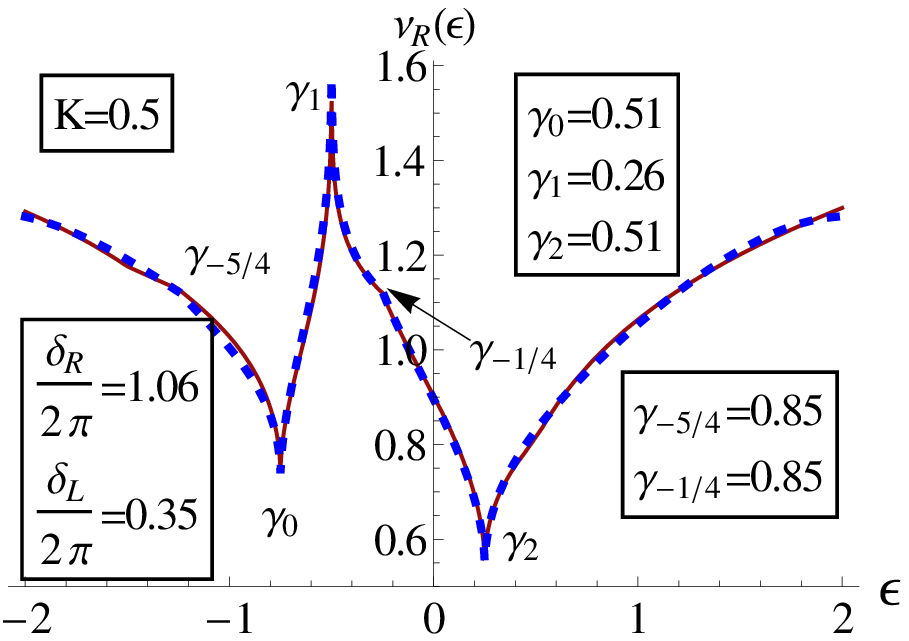}
\includegraphics[width=210pt]{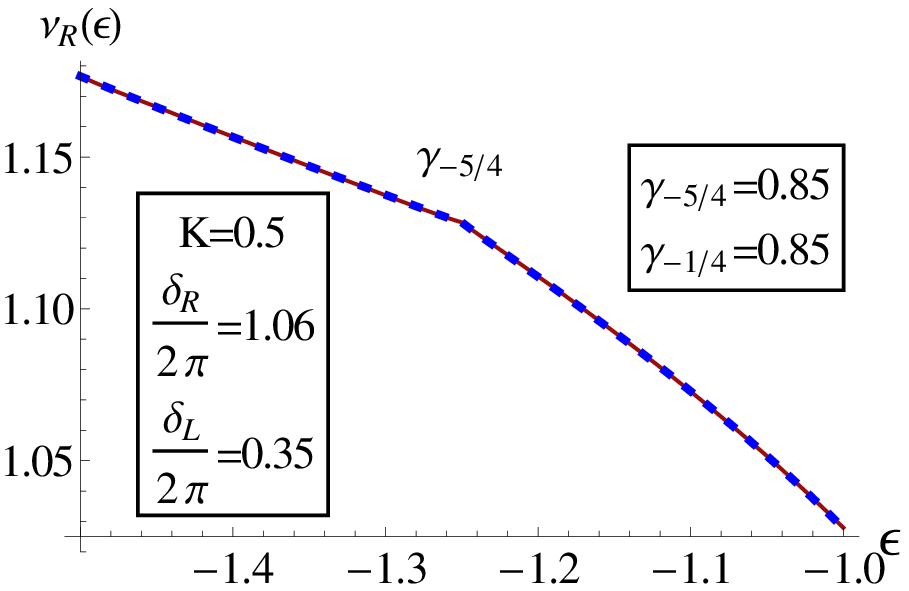}
\includegraphics[width=210pt]{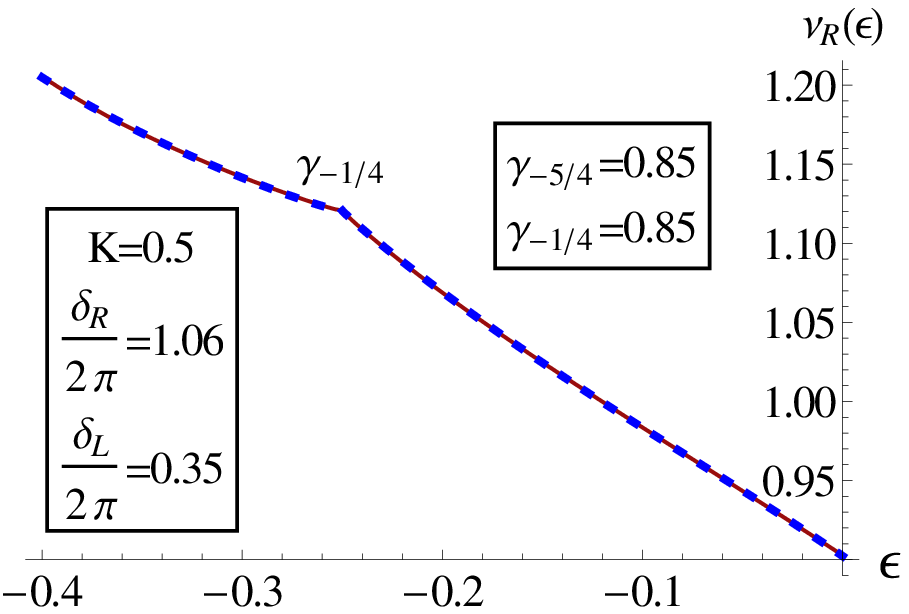}
\caption{\small The density of states in the Luttinger liquid
  coupled to two reservoirs with  
the triple-step distribution $n_t(\epsilon)$. There is no dephasing,
so that the singularities are sharp.  
At weak interaction ($K=0.7$, upper right plot), 
the most important terms in the right determinant
$\overline{\Delta}_{R\tau}[\delta_R]$ oscillate as a function of time  
$\tau$ with frequencies $\epsilon_k$, $k=0\,,\, 1\,,\, 2$. On the
other hand,  the
leading contribution to the left determinant decays without
oscillations, and the oscillatory terms are very small (decay with
much larger exponents).
This leads to the singular behavior of the density of states at
$\epsilon_k$. As the Luttinger parameter decreases ($K=0.5$, upper
left plot),  
the subleading oscillating terms in $\overline{\Delta}_{L\tau}[\delta_L]$
come into play. This leads to additional (weak)  
singularities in $\nu_R(\epsilon)$ at $\epsilon=-5/4$ and
$\epsilon=-1/4$. The corresponding regions of energy are magnified in
the lower plots.
}
\label{Fig:LuttingerLiquidInversePopulationInversePopulation}
\end{figure*}

Let us now turn to another application of Toeplitz determinants, the
tunneling into the Luttinger liquid. We begin by considering the
simplest case, when the incoming right-moving
electrons have the double-step distribution function 
$n_d(\epsilon)$, while the left movers are held at zero temperature. In
this case the determinant $\overline{\Delta}_{L\tau}[\delta_L]$ in
Eq.~(\ref{d3}) is  
identically equal to unity. If the  interaction is not too strong and
one is interested in the density of states for the left-movers, the
phase $\delta_R$ entering the non-trivial determinant  
$\overline{\Delta}_{R\tau} [\delta_R]$ is close to zero. On the other
hand, the correlation functions of the right-movers are given by the
determinants at phase $\delta_R$ close to  $2\pi$.  
Correspondingly, the dominant singularity in the density of states   
$\nu_L(\epsilon)$  for the left particles is the one at $\epsilon=0$ while 
main singularities of $\nu_R(\epsilon)$  are at 
$\epsilon=\epsilon_0\,,\, \epsilon_1$. This behavior is illustrated in
Fig. \ref{Fig:LuttingerLiquidDoubleStepEquilibriumLeftRight}. 
Note that the
left-moving electrons are dephased much stronger\cite{GGM_long2010} 
than the right-moving.

The behavior of $\nu_L(\epsilon)$ at stronger interaction (see
Fig. \ref{Fig:LuttingerLiquidDoubleStepEquilibriumLeft}) 
demonstrates the non-monotonous dependence of the dephasing on the
Luttinger liquid parameter $K$.  For  
$K<(3-\sqrt{5})/2\approx 0.38$, the phase $\delta_R>\pi$, and the
leading singularities in $\nu_L$ are those at $\epsilon_0$ and
$\epsilon_1$. They can be clearly seen if one plots the second
derivative of the density of states with respect to energy
(Fig. \ref{Fig:LuttingerLiquidDoubleStepEquilibriumLeft}, left
panel). Note that the smearing of those singularities {\it decreases}
(i.e. singularities sharpen) with {\it increasing} interaction
strength $K = 0.3 \to 0.25 \to 0.2$, as $K$
evolves towards $K=3-2\sqrt{2}\approx 0.17$, where $\delta_R=2\pi$ and
the dephasing is absent.  

Finally, we consider an interacting wire 
with triple-step distribution $n_t$ for both  left and right moving electrons. 
In this case, both determinants in Eq.~(\ref{d3}) are nontrivial. 
The corresponding density of states is shown in
Fig. \ref{Fig:LuttingerLiquidInversePopulationInversePopulation}.  
At weak interaction ($K=0.7$, upper-left panel of
Fig. \ref{Fig:LuttingerLiquidInversePopulationInversePopulation}), the
right determinant $\overline{\Delta}_{R\tau}[\delta_R]$ oscillates as
a function of time  
$\tau$ with frequencies $\epsilon_k$, $k=0\,,\, 1\,,\, 2$, while the
left determinant decays mostly without oscillations, 
This leads to the singular behavior of the density of states at $\epsilon_k$. 
As the Luttinger parameter decreases ($K=0.5$, upper-right panel of
Fig.\ref{Fig:LuttingerLiquidInversePopulationInversePopulation}),   
sub-leading oscillating terms in $\overline{\Delta}_L\tau[\delta_L]$
come into play and additional (weak) singularities in
$\nu_R(\epsilon)$ appear at $\epsilon=-5/4$ and $\epsilon=-1/4$. 


 \section{Summary and outlook}
\label{Sec:Conclusions}

To summarize, we have explored single-particle Green functions were   
of many-body fermionic systems in non-equilibrium settings
characterized by multiple-step energy distribution
functions. 
By using  a periodic  ultraviolet regularization, the problem is
reduced to that of Toeplitz determinants. We have carried out  
numerical calculation of the corresponding Toeplitz determinants and
thus obtained the results for the non-equilibrium Green functions 
in the entire energy range. Further,
by employing  the  extended Fisher-Hartwig conjecture, we have analytically  
determined  the  energy dependence of the Green functions 
near each of the  Fermi edges.

The obtained Green functions show, in the energy representation, 
power-law singularities near multiple edges. The singularities are in
general characterized by different power-law exponents and are  
smeared by dephasing processes. 
In  the special case of a distribution function with population
inversion that alternates between $n=1$ and $n=0$, the dephasing is
absent (i.e. the singularities are sharp) and the TDOS (or the absorption
rate) exhibits enhancement and suppression  in alternating succession.

The results of the numerical and analytical methods  perfectly agree,  
thus confirming the validity of the extended Fisher-Hartwig conjecture.

We close the paper by listing some of future research directions:

\begin{itemize}

\item It would be interesting to see whether an explicit form of
  correction terms within each harmonic [those abbreviated by
  $+\ldots$ in Eq.~ (\ref{Delta_asymptotic})] can be found.   
Further, a rigorous mathematical proof of the extended Fisher-Hartwig
  conjecture would be certainly desirable. 

\item One can consider many-body correlation functions in the
  non-equilibrium setups discussed above. This problem can be reduced
  to determinants that are of a form more general than the Toeplitz
  one. Some results in this direction will be reported soon \cite{in-preparation}.

\item It would be important to further extend the Fisher-Hartwig
  conjecture in order to include Toeplitz matrices with matrix
  symbols. 

\end{itemize}

\section{Acknowledgments}

This paper has been prepared for publication in a volume commemorating
Yehoshua Levinson. Two of us were very fortunate to know Yehoshua
personally, to attend his lecture courses, to collaborate with
him, and to enjoy numerous physics discussions with him. We owe to him
much of our understanding of non-equilibrium phenomena in
condensed-matter physics.

We thank A.R.~Its for illuminating discussions of the Fisher-Hartwig
conjecture \cite{deift09} and its extended version \cite{Gutman10}.
We also thank R. Berkovits for useful discussions and collaboration on
an early stage of this work, and V.V.~Cheianov and D.A.~Ivanov for
useful discussions. I.V.P. 
acknowledges financial support by  Alexander von Humboldt foundation. 
Financial support by German-Israeli Foundation, Israeli Science
Foundation, DFG Center for Functional Nanostructures, and
Bundesministerium f\"ur Bildung und Forschung is gratefully
acknowledged. A.D.M. thanks KITP UCSB for hospitality during the
completion of this work.

\end{document}